\documentclass[twocolumn,showpacs,amsmath,amssymb,pre,superscriptaddress]{revtex4}

\usepackage{graphicx}% Include figure files
\usepackage{amsmath,amssymb}
\usepackage{dcolumn}% Align table columns on decimal point
\usepackage{bm}% bold math
\usepackage{subfigure} %bilder nebeneinander
\usepackage{natbib}
\usepackage[ansinew]{inputenc}

\begin{document}

\title{Dynamical trapping and chaotic scattering of the harmonically driven barrier}

\date{\today}


\author{Florian R. N. Koch}
\email[]{florian.koch@physi.uni-heidelberg.de}
\affiliation{Physikalisches Institut, Universit\"at Heidelberg, Philosophenweg 12, 69120 Heidelberg, Germany}

\author{Florian Lenz}
\email[]{lenz@physi.uni-heidelberg.de}
\affiliation{Physikalisches Institut, Universit\"at Heidelberg, Philosophenweg 12, 69120 Heidelberg, Germany}

\author{Christoph Petri}
\email[]{petri@physi.uni-heidelberg.de}
\affiliation{Physikalisches Institut, Universit\"at Heidelberg, Philosophenweg 12, 69120 Heidelberg, Germany}

\author{Fotis K. Diakonos}
\email[]{fdiakono@phys.uoa.gr}
\affiliation{Department of Physics, University of Athens, GR-15771 Athens, Greece}

\author{Peter Schmelcher}
\email[]{Peter.Schmelcher@pci.uni-heidelberg.de}
\affiliation{Physikalisches Institut, Universit\"at Heidelberg, Philosophenweg 12, 69120 Heidelberg, Germany}
\affiliation{Theoretische Chemie, Institut f\"ur Physikalische Chemie, Universit\"at Heidelberg, INF 229, 69120 Heidelberg, Germany}

\begin{abstract} \label{sec_abstract}
A detailed analysis of the classical nonlinear dynamics of a single driven square potential barrier with harmonically oscillating position is performed. The system exhibits dynamical trapping which is associated with the existence of a stable island in phase space. Due to the unstable periodic orbits of the KAM-structure, the driven barrier is a chaotic scatterer and shows stickiness of scattering trajectories in the vicinity of the stable island. The transmission function of a suitably prepared ensemble yields results which are very similar to tunneling resonances in the quantum mechanical regime. However, the origin of these resonances is different in the classical regime.
\end{abstract}

\maketitle

\section{Introduction}

Harmonically driven potentials appear in many areas of modern physics, particularly in mesoscopic electronic semiconductor devices and other micro- and nano-structures driven by external voltages or applied laser fields. They also play a role for ultra-cold atomic wave packets exposed to optical barriers and photo-induced dynamics in strong laser fields or dissociation processes of molecules on solid surfaces. The strong external driving of the system typically leads to nonlinear quantum effects and chaos in the corresponding classical systems. Two archetypical potentials have been investigated in detail in the literature, the driven potential well and the driven potential barrier. The periodic driving can be either a driving of the height or of the position of the potential. In the following, we will give a short overview of the known features of these systems.

An early study of a vertically oscillating rectangular potential barrier, i.e. a potential with harmonically oscillating height, in \cite{Landauer1982} aimed to derive an expression for the tunneling time through potential barriers. Particles interacting with a driven potential barrier can absorb or emit quanta of \(\hbar \omega\), where \(\omega\) is the driving frequency. This leads to frequency-dependent resonances in the tunneling probability through the vertically oscillating barrier, as shown in \cite{Hagmann1995} for a rectangular barrier and a raised cosine potential. These resonances can, according to \cite{Lefebvre2004}, be interpreted as poles of the scattering amplitudes in the complex plane. If the potential barrier is delta-shaped, it possesses a set of leaky bound states, which have been detected in \cite{Bagwell1992} by locating the complex energy poles of the transmission amplitude. These semi-bound states lead to additional resonances in the transmission function. Other works analyzed the tunneling through a vertically driven Gaussian-shaped barrier \cite{Pimpale1991} and through the ``Eckart''-barrier \cite{Ge1996}. Both found an amplification of tunneling for intermediate frequencies, but did not detect a resonant behavior, which is due to the parameters chosen in these works, according to \cite{Chiofalo2003}. A vertically oscillating rectangular barrier, enclosed in a rigid potential box, has been studied in the classical regime, see refs.~\cite{Mateos1998,Mateos1999,Leonel2004}. The phase space of this system is mixed, with a purely chaotic layer at low energies, KAM-islands at intermediate energies and invariant spanning curves at high energies. The distribution of transversal times through the oscillating barrier is asymptotically algebraic with an exponent \(\gamma=-3\). The Lyapunov exponent in the chaotic sea was also calculated for this system in \cite{Leonel2004} as a function of the parameters. The Lyapunov exponent changes abruptly whenever an invariant spanning curve, separating different parts of the chaotic sea, is destroyed. If the driving is stochastic instead of harmonic, the system has no invariant spanning curve for high energies and the particles exhibit normal Fermi-acceleration (\cite{Leonel2004}).

A laterally oscillating potential is often the result of a ``Kramers - Henneberger'' transformation of an ac-driven static potential. For high driving frequencies the tunneling probability through an ac-driven rectangular barrier shows resonances at small energies for which the static barrier would be entirely intransparent, see \cite{Vorobeichik1998} and \cite{Chiofalo2003}. This can be described as resonant tunneling into quasi-stable bound states of the effective time-averaged potential, which has a double-barrier structure. For intermediate frequencies, the scattering process is dominated by inelastic processes and strong sidebands in the energy spectrum \cite{Chiofalo2003}. Such a resonant behavior was not found in the ac-driven Gaussian-shaped barrier \cite{Pimpale1991} and the ``Eckart''-barrier \cite{Ge1996}. Instead, these two systems exhibit phase-sensitive tunneling resonances for intermediate frequencies, which can be explained by an increase in the relative kinetic energy of incoming particles when the barrier approaches them. A moving potential barrier can also be used to tailor wave packets or to split an initial pulse into several well-separated coherent pulses, see \cite{Papachristou2007,Kuhn2005}.

Potential wells with oscillating bottom have been studied in \cite{Henseler2000} and \cite{Henseler2001}. In the classical regime, the oscillating square well is pseudo-integrable and therefore not a chaotic scatterer (see also \cite{Luna-Acosta2001}), whereas the oscillating smooth well has a stable KAM-island in phase space which leads to chaotic scattering. In the quantum mechanical regime both the smooth and the rectangular oscillating well accommodate quasi-bound Floquet states. Multi-photon processes couple incoming particles to these states, leading to resonances, which are visible as dips in the transmission function and as peaks in the dwelltime.

The tunneling through the laterally oscillating square well has resonances due to quasi-bound Floquet states as well (\cite{Henseler2000} and \cite{Henseler2001}). However, due to the lateral driving these states belong to an effective double-well potential which leads to the formation of tunneling doublets. In the classical regime, the ac-driven square well has a stable KAM-island in phase space which leads to chaotic scattering. A signature of this KAM-island is visible in representations of scattering wave functions in terms of Wigner functions. The avoided crossings of the Floquet quasi-energies of that system are studied in \cite{Timberlake1999}. Sharp crossings of quasi-energies lead essentially only to a relabeling of the states, whereas broad crossings, in which more than two states take part, completely alter the Floquet states and increase the high harmonic generation.

Many other elementary systems have been studied as well: The so called tunneling diode, which consists of an oscillating quantum well between two static barriers, is investigated in \cite{Wagner1994} and \cite{Wagner1995}. In this system the driving creates additional sidebands due to the absorption or emission of oscillation quanta \(\hbar \omega\). The transmission through all sidebands exhibits a strong quenching at certain parameters. The classical phase space of an infinite array of vertically oscillating potential wells is mixed chaotic at small energies and has regular spanning curves at high energies, see \cite{Luna-Acosta2001}. The motion of a particle in such a system resembles closely a random walk. Driven double-wells have been studied in \cite{Grossmann1991} and \cite{Vorobeichik99}. The periodic driving can be used in these systems to entirely suppress the tunneling between the two wells, if the driving frequency is tuned to an exact crossing of the Floquet energies of the ground state doublet of the unperturbed double-well.

Although several works deal with the tunneling through a laterally driven square potential, the classical dynamics in this system is largely unknown. The transmission of classical particles through a laterally driven Gaussian-shaped potential has already been calculated in \cite{Pimpale1991}. However, this work does neither cover the whole range of possible parameters nor does it give any clue about the phase space structure, which is essential for the scattering process. The aim of this work is to close this gap and to provide a comprehensive survey of the periodically driven square barrier with oscillating position in the classical regime. We will analyze the entire phase space, where we find stable KAM islands, understand the underlying dynamics and the chaotic scattering process. We will also explore the full range of possible system parameters and make comparisons with quantum mechanical results for the transmission probability.

This work is organized as follows: In section \ref{sec_mapping} we introduce the system of the driven barrier and derive a mapping to describe the dynamics. In section \ref{sec_phasespace} we analyze the phase space where we find a stable island of quasi-periodic orbits. The position, size and parameter dependence of the stable island is studied in detail. The scattering dynamics, particularly the influence of the stable manifolds of the unstable periodic orbits, are the subject of section \ref{sec_scattering}. Due to the existence of the stable island, the scattering dynamics is chaotic and trapping in the sticky region of phase space is possible although the barrier is purely repulsive. Comparisons of the classical transmission probability with the quantum mechanical tunneling probability are made in subsection \ref{subsec_QM}. Finally, a summary is given in the last section \ref{sec_summary}.

\section{The driven barrier and its mapping} \label{sec_mapping}

\begin{figure}[htbp]
\includegraphics[width=0.5 \textwidth]{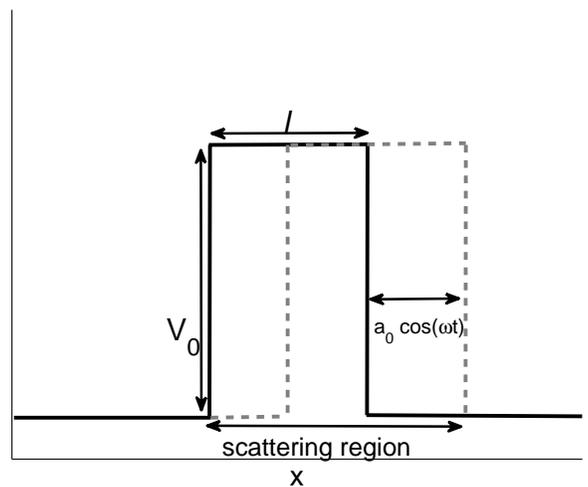}
\caption{The ac-driven potential barrier.}
\label{fig_model}
\end{figure}

Our classical system consists of a one-dimensional laterally oscillating potential of a finite and constant height \(V_0\) and width \(l\), see Fig.~ \ref{fig_model}. The driving function is assumed to be harmonic, with a driving amplitude \(a_0\) and frequency \(\omega\)
\begin{equation}
V(x,t)=V_{0} \Theta(\frac{l}{2}- |x- a_{0}\cos(\omega t)| )
\label{equ_V(x,t)}
\end{equation}

Although the dynamics of the system is continuous, the forces acting on the particle are point-like and the particles move ballistically between collisions with either of the edges of the barrier. Therefore, it is sufficient to describe the dynamics in terms of a discrete mapping between collisions. To describe the particle-barrier interaction, we transform all coordinates into the frame of reference of the barrier, to the variables \(\tilde{x}\) and \(\tilde{v}\). Even though this coordinate frame is accelerated and, therefore, not an inertial frame, momentum and energy are conserved for the infinitesimally small time span of the interaction.
\begin{equation}
x\rightarrow \tilde{x}=x-a_{0}\cos(\omega t) \qquad
v\rightarrow \tilde{v}=v+a_{0} \omega \sin(\omega t)
\end{equation}
When colliding with the barrier from the \emph{outside} a particle of mass \(m\) is transmitted \emph{into} the barrier if its
kinetic energy, relative to the barrier, surpasses the barrier height \(V_0\).
\begin{equation}
E_{kin}=\frac{m}{2}\tilde{v}_{n}^{2}\geqslant V_{0}
\end{equation}
In this case the particle is decelerated due to energy conservation
\begin{equation}
\tilde{v}_{n+1}=\sqrt{\frac{2}{m}(E_{kin}-V_{0})} .
\end{equation}
If it is not transmitted, it is reflected and its new velocity becomes
\begin{equation}
\tilde{v}_{n+1}=-\tilde{v}_{n} .
\end{equation}
Likewise, if a particle hits the barrier coming from its \emph{inside}, it is always transmitted and accelerated
\begin{equation}
\tilde{v}_{n+1}=\sqrt{\frac{2}{m}(E_{kin}+V_{0})} .
\end{equation}
Transforming these equations back to the laboratory frame yields the equations of motion
\begin{equation}
\begin{split}
v_{n+1}=v_{b}(t_{n+1})+ \text{sign}(v_{n}-v_{b}(t_{n+1})) \\ \sqrt{(v_{n}-v_{b}(t_{n+1}))^{2}\pm\frac{2}{m} V_{0}} ,
\label{equ_vn+1_a}
\end{split}
\end{equation}
\begin{center}
for transmission and
\end{center}
\begin{equation}
v_{n+1}=2 v_{b}-v_{n} ,
\label{equ_vn+1_b}
\end{equation}
if the particle is reflected, where \(v_{b}(t)=-a_0 \omega \sin(\omega t_{n+1})\) is the barrier's velocity at the time of the collision \(t_{n+1}\) and the sign \(\pm\) depends on whether the particle is transmitted into the barrier (\(-\)) or leaves the barrier (\(+\)).
The time \(t_n\) is mapped on the time \(t_{n+1}\) of the next collision of the particle with one of the barrier's edges. Therefore \(t_{n+1}\) is the smallest solution of
\begin{equation}
x_b(t_{n+1})=x_n+v_n(t_{n+1}-t_n) , \label{equ_coll}
\end{equation}
where \(x_{b}\) can be either edge of the barrier, \(x_b=a_0\cos(\omega t)\) or \(x_b=a_0\cos(\omega t)+l\). If equation (\ref{equ_coll}) has no solutions for \(t_{n+1}>t_n\) then the particle does not collide with the barrier again and escapes. This implicit equation can be solved only numerically. It is important to make sure that the numerically calculated collision time is the \emph{smallest} solution of equation (\ref{equ_coll}), because many effects, like stickiness in phase space (see sections \ref{sec_phasespace} and \ref{sec_scattering}), are susceptible to errors in the time mapping.

The mapping shows that the five parameters of the system, the barrier's height \(V_{0}\) and thickness \(l\), the driving frequency \(\omega\), the amplitude \(a_{0}\) and the particle's mass \(m\), can be reduced to just two effective parameters by an appropriate scaling transformation. Equations (\ref{equ_vn+1_a}) and (\ref{equ_vn+1_b}) then become:
\begin{align}
\label{equ_coord_trafo}
x & \rightarrow x'=\frac{x}{a_{0}} \qquad t \rightarrow t'=t \omega   \qquad v  \rightarrow v'=\frac{v}{\omega a_{0}}    \\
\begin{split}
v'_{n+1} = -\sin(t_{n+1}')+ \text{sign}\left(v'_{n}+ \sin(t_{n+1}')\right) \\ \sqrt{\left(v'_{n}+\sin(t_{n+1}')\right)^{2}\pm\frac{V_{0}}{V_{\omega}}}
\end{split} \\
&v'_{n+1} = -2 \sin(t_{n+1}') -v'_{n}
\end{align}
The only parameter left in the mapping is \( \frac{V_{0}}{V_{\omega}} \), where \(V_{\omega}=\frac{m}{2} a_{0}^{2} \omega^{2}\) is the maximum kinetic energy a particle, which is at rest in the laboratory frame, can have in the barrier's frame of reference. The second parameter, \(\frac{l}{a_{0}} \), is the barrier's thickness measured in units of the amplitude and appears in the implicit equation (\ref{equ_coll}) for the time mapping. In the following, we will scale the energy in units of \(V_{\omega}\) and use the scaled coordinates in all calculations while keeping the same notation. The system is therefore completely described by the two parameters \(V_0\) and \(l\).

\section{Phase space structure} \label{sec_phasespace}

We visualize the structures in phase space in terms of Poincar\'e surface of sections by mapping all collisions of the particle with either side of the barrier to the Poincar\'e section. This is equivalent to mapping all intersections of the trajectories in the 3-dimensional phase space with the two-dimensional manifold \(\Omega\) defined by the barrier's motion
\begin{equation*}
\Omega=\left\{ \left( \begin{array}{c} t \\ x_{b}(t) \\ v  \end{array}    \right)   \qquad | t,v \in \Re \right\}
\end{equation*}
where \(x_b(t)\) is the position of either of the barrier's edges.  Since the driving function is assumed to be periodic, we can use the phase \(\varphi\), defined as \( \omega t \mod 2\pi \), instead of the time as a coordinate. The mapping naturally operates only on the manifold defined by \(\Omega\), because it always maps a point in phase space on the point of the next collision with the barrier. The Poincar\'e section is made unique by mapping only intersections with one chosen edge of the barrier in a specified direction, i.e. \(v>v_b\) or \(v<v_b\), instead of all collisions. Thus the position and phase \(\varphi\) are uniquely connected by the driving function and one of the two coordinates becomes redundant. In the following, we will discuss Poincar\'e sections in which we plot the particle's velocity after a collision over the phase \(\varphi\) of the oscillating barrier.

\begin{figure}[htbp]
\subfigure[]{
\includegraphics[width=0.48 \textwidth]{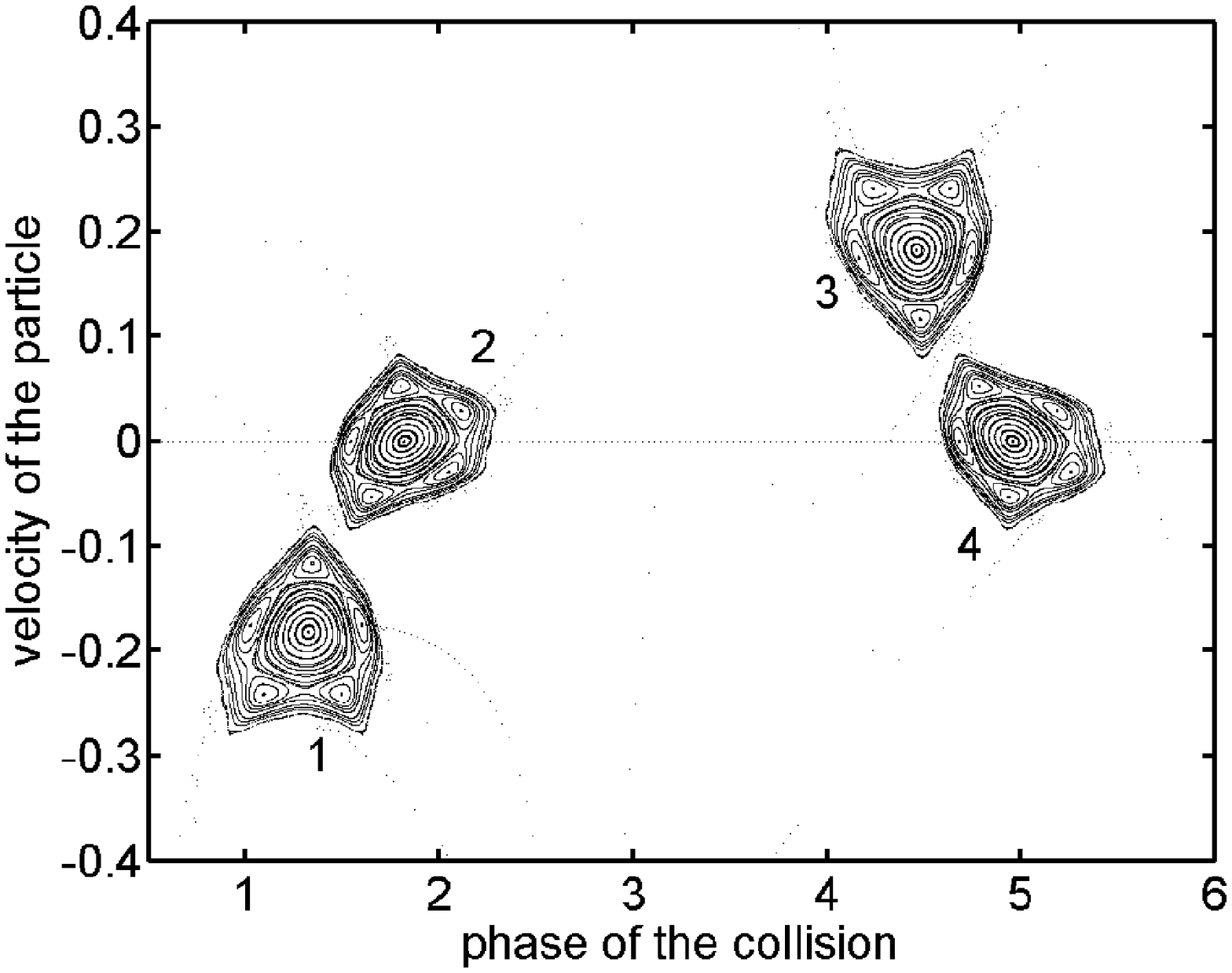}
\label{fig_poincare}
}
\subfigure[]{
\includegraphics[width=0.48 \textwidth]{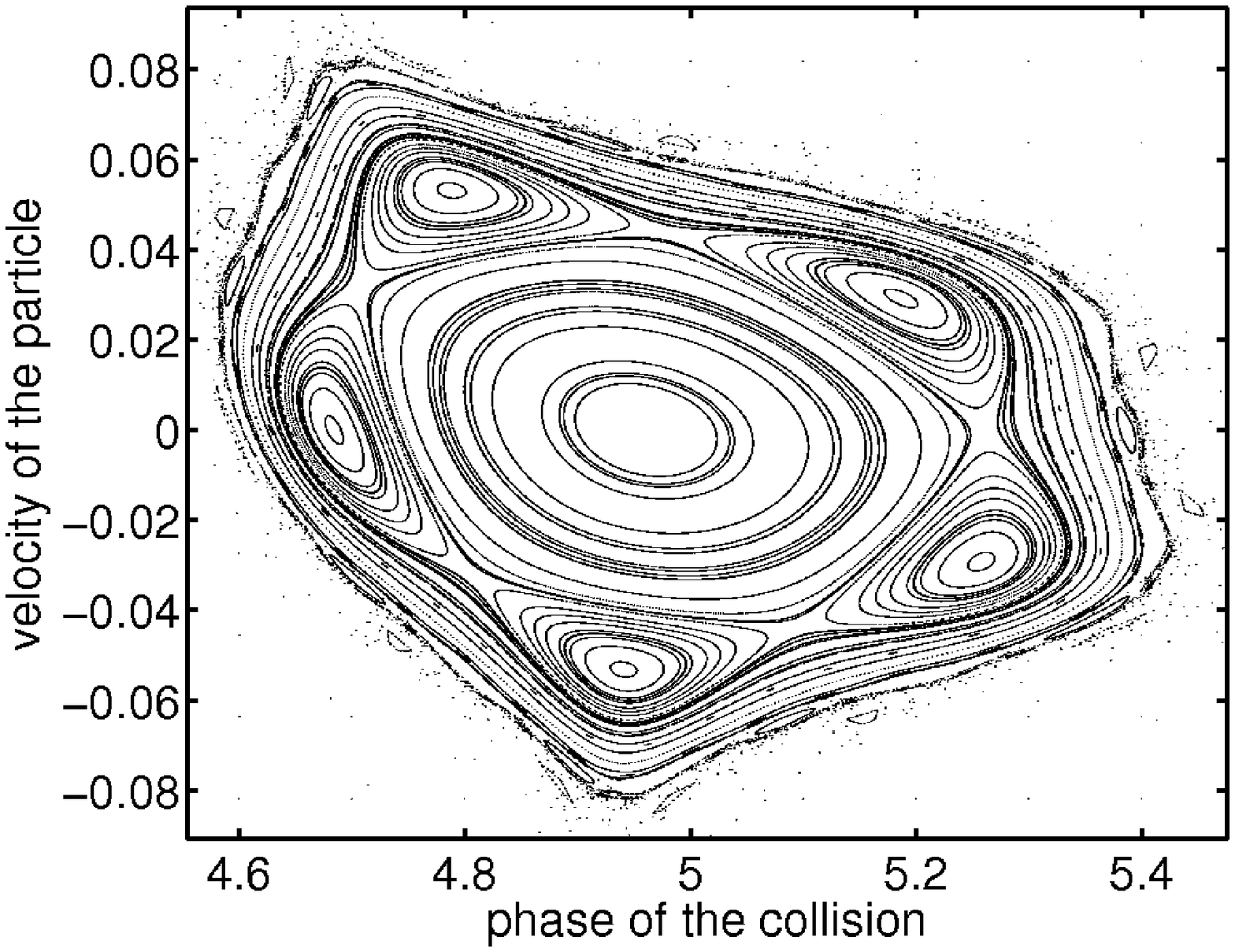}
\label{fig_poincare_zoom}
}
\caption{(a): A typical Poincar\'e section showing trapped particles (b): Enlargement of island 4.}
\end{figure}

We covered the entire phase space with a fine grid of initial conditions to guarantee that all relevant structures are being shown in our Poincar\'e sections. The resulting Poincar\'e section is plotted in Fig.~\ref{fig_poincare}.
The system's parameters are: \(V_0=0.32\) and \(l=0.4\). These parameters are typical for an experimental setup using semiconductor structures driven by external voltages or applied Laser fields, \cite{Hanson2007,Davies1998}.

The phase space of the ac-driven barrier has four stable KAM-islands whose center is a stable periodic orbit of period four. This means that, through the driving, the repulsive potential can trap particles in a small part of phase space. This kind of dynamical trapping works only for a harmonic driving law or other similarly curved functions. The periodic orbit at the center of the island is stable because it lies symmetrically around the inflection points of the harmonic driving function. A sawtooth shaped driving function, for example, does not create stable orbits. \footnote{That a sawtooth shaped driving law does not lead to a stable periodic orbit can be seen directly from equ.~(\ref{equ_vn+1_a}): When a particle passes through the barrier driven by a sawtooth shaped driving law, the velocity of the barrier is the same at the collision with the left and right edge, unless the collision takes place around the turning point of the barrier. Therefore, we see from equ.~(\ref{equ_vn+1_a}) that the transmission through the barrier does not change the velocity of the particle, which forbids the creation of a periodic orbit.} An enlargement of the fourth island in Fig.~\ref{fig_poincare} is plotted in Fig.~\ref{fig_poincare_zoom} and shows the typical structure of an elliptic fixed point surrounded by a stable island of quasi-periodic orbits and chains of sub-islands, which are the remnant of dissolved quasi-periodic orbits. The thin transition zone at the edge of the island is mixed chaotic and contains a fractal structure of sub-islands. The space outside the stable orbits contains only few points because this part of phase space is visited only by trajectories which leave the open system after a few collisions, whereas those on regular orbits stay in the scattering region (defined as the space that is covered by the barrier's oscillation) indefinitely. The Poincar\'e section in Fig.~\ref{fig_poincare} is \emph{not} unique because it shows all collisions, with both edges of the barrier and in both directions. Structure 1 and 4 correspond to collisions with the left edge, structure 2 and 3 correspond to collisions with the right edge.
The four regular islands are symmetrical, the first and second structures are identical to the third and fourth with their phases increased by \(\pi\) and the sign of their velocity inverted. This reflects the symmetric properties of the driving function \(a_{0}\cos(\omega t)\): \(\cos(\pi-\varphi)=\cos(\pi+\varphi)\) and \(\cos(\pi/2-\varphi)=-\cos(\pi/2+\varphi)\)

\begin{figure}[htbp]
\subfigure[]{\includegraphics[width=0.48 \textwidth]{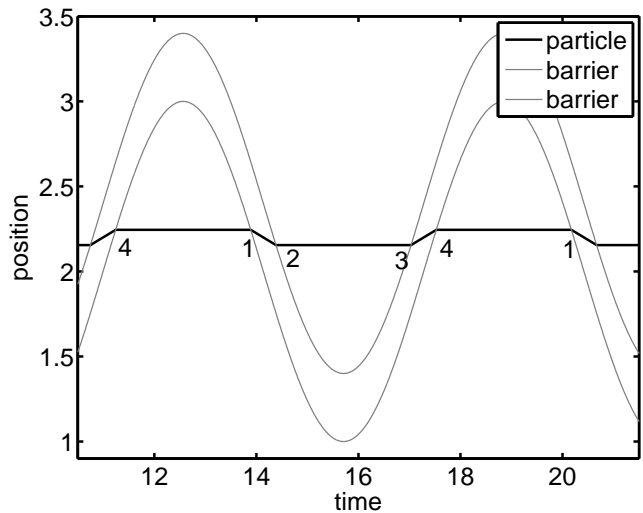} \label{fig_xt-fixpoint}}
\subfigure[]{\includegraphics[width=0.48 \textwidth]{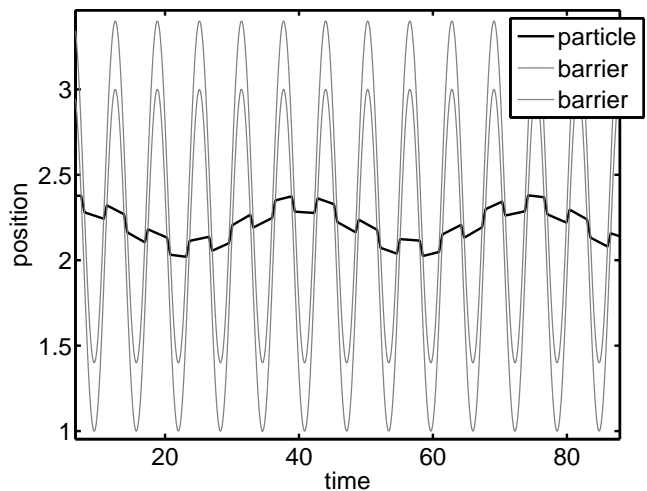} \label{plot_xt-trapped-example}}
\caption{Motion of a trapped particles. The black line represents the particle, the two gray lines represent the edges of the barrier. (a) is the trajectory associated with the central periodic orbit of the stable island, (b) is a typical trajectory of a quasi-periodic orbit. The numbers in (a) correspond to the island numbers in Fig.~\ref{fig_poincare}.}
\end{figure}

The trajectory of the central periodic orbit of the stable island, plotted in Fig.~\ref{fig_xt-fixpoint}, is as following: Starting left of the barrier at (4) with a velocity of zero, the particle is hit by the barrier at (1). Due to the relatively high negative velocity of the barrier at \(\varphi_1\), the particle is transmitted into the barrier and accelerated in negative direction. This creates the stable fixed point of the island 1 in Fig.~\ref{fig_poincare} and the corresponding stable island 1 if we vary the initial conditions in the neighborhood of the fixed point. The barrier then overtakes the particle inside of it and the particle collides with the right edge of the barrier at (2). Since the velocity of the barrier at the points (1) and (2) is identical (\(v_b(\varphi_1)=v_b(\varphi_2)\)), the velocity of the particle becomes zero again. This collision corresponds to island 2 in Fig.~\ref{fig_poincare}. After the barrier has reached its minimum position and turning point at \(\varphi = \pi\), it moves back in positive direction and hits the particle at (3). This collision accelerates the particle in positive direction and the particle moves with the barrier until it is overtaken by it at (4), where the velocity of the particle becomes zero again and the periodic cycle starts again. Due to the symmetry of the driving function (\(\cos(\varphi)\)), the collisions (1) and (2) are symmetrical to (3) and (4).

The trajectories of the quasi-periodic orbits can be understood as a perturbation of the periodic orbit described above. A typical trajectory of such an orbit is plotted in Fig.~\ref{plot_xt-trapped-example} and shows both modes of the motion: the periodic hopping between the four islands inherited form the stable periodic orbit and an overlaid harmonic oscillation. This sinusoidal motion is represented as closed orbit in the Poincar\'e section. The frequency of this harmonic oscillation, measured by a Fourier transformation, is equal to the frequency by which the particles rotate on the quasi-periodic orbit around the elliptic fixed points in the Poincar\'e section.

It is possible to calculate the position of the central elliptic fixed points, here we will do it for island number 1. As Fig.~\ref{fig_xt-fixpoint} shows, the velocity of the periodic orbit is zero while the particle is outside of the barrier in structure 2 and 4 (\(v_0=v_2=0\)). Due to the symmetry of the system, the collision points are symmetric around \(\frac{\pi}{2}\) in their phases \(\varphi_1\) and \(\varphi_2\) (\(\varphi_2=\pi-\varphi_1\)) and around the equilibrium position in their positions \(x_1\) and \(x_2\). Therefore, the velocity in structure 1 has to be:

\begin{equation}
v_1=\frac{\Delta x}{\Delta t}=- \frac{x_{b}(\varphi_1)-\frac{l}{2}}{(\frac{\pi}{2}-\varphi_1)} \qquad
x_{b}(\varphi_1)=\cos(\varphi_1)
\label{equ_fix_v1}
\end{equation}
where \(x_b\) is the position of the left edge of the barrier. This velocity can be calculated from equation (\ref{equ_vn+1_a}):
\begin{equation}
v_1=v_{b}(\varphi_1)+ \text{sign}\left(v_0-v_{b}(\varphi_1)\right) \sqrt{\left(v_0-v_{b}(\varphi_1)\right)^{2} - V_{0}}
\label{equ_fix_v2}
\end{equation}
The initial velocity \(v_0\) is zero and \(v_{b}=-\sin(\varphi)\). The result is an implicit equation for \(\varphi_1\):
\begin{equation}
\begin{split}
f(\varphi_1):&= \left(\frac{2 \cos(\varphi_1) - l}{\pi-2\varphi_1} \right)^2 \left(\frac{2\sin(\varphi_1)(\pi-2\varphi_1)}{2\cos(\varphi_1) - l} -1\right) \\
&= V_0
\end{split}
\label{equ_fix}
\end{equation}

\begin{figure}[htbp]
\includegraphics[width=0.48\textwidth]{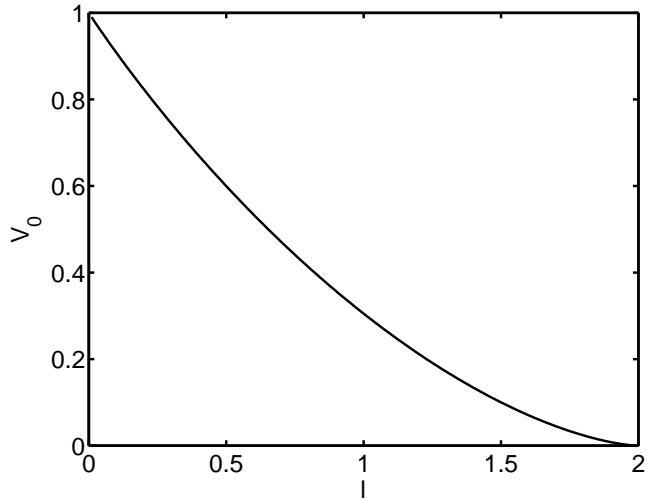}
\caption{Maximum barrier potential allowing for periodic orbits as a function of the barrier's thickness. Only pairs of parameters below this curve lead to stable orbits.}
\label{fig_fix2}
\end{figure}

This implicit equation determines the phase of the first elliptic point. For values of \(l \in [0,2]\), the equation \(f(\varphi)=V_0\) has two solutions in the interval \([0,\frac{\pi}{2}]\). The physically relevant solution lies to the right of the function's maximum. For \(V_0 > f_{max} \), where \(f_{max}\) is the maximum value of \(f\) at a chosen value of \(l\), the implicit equation (\ref{equ_fix}) has no solutions. This means that the barrier's potential is too strong to allow for trapped orbits. The function has no positive values for \(l\geq2\) and the second root is greater than \(\frac{\pi}{2}\) for \(l=0\). The elliptic orbits disappear in both cases. Figure \ref{fig_fix2} shows the maximal value of \(f(\varphi)\) as a function of \(l\). Only pairs of parameters \((V_0,l)\) \emph{below} the curve allow for trapping of particles. For sets of parameters \((V_0,l)\) above the curve in Fig.~\ref{fig_fix2} the phase space contains no bound orbits. To be precise, the central periodic orbit does not just become unstable for other parameters. The periodic orbit, the surrounding elliptic island and all unstable periodic orbits cease to exist!

The maximal values of \(V_0=1\) and \(l=2\) can be easily understood physically: \(V_{\omega}\) is the maximum kinetic energy a particle can have in the barrier's frame of reference if it is at rest in the laboratory frame. If \(V_{0}\) is greater than \(V_{\omega} \), then particles with velocity zero will never be transmitted into the barrier. Since all stable orbits cross the \(v=0\) axis, this would destroy all of them. For \(l>2\), \(v_1\) in equation (\ref{equ_fix_v1}) would have to be positive, which is physically impossible (see discussion of the periodic orbit, Fig.~\ref{fig_xt-fixpoint}).

The phases of the other fixed points can be calculated from the phase and position position of the fixed point of island 1 by using the symmetry properties of the driving function:
\begin{align*}
\varphi_{2} &= \pi - \varphi_{1} \\
\varphi_{3} &= \pi + \varphi_{1} \\
\varphi_{4} &= 2\pi - \varphi_{1}
\end{align*}

\begin{figure}[htbp]
\subfigure[]{\includegraphics[width=0.22\textwidth]{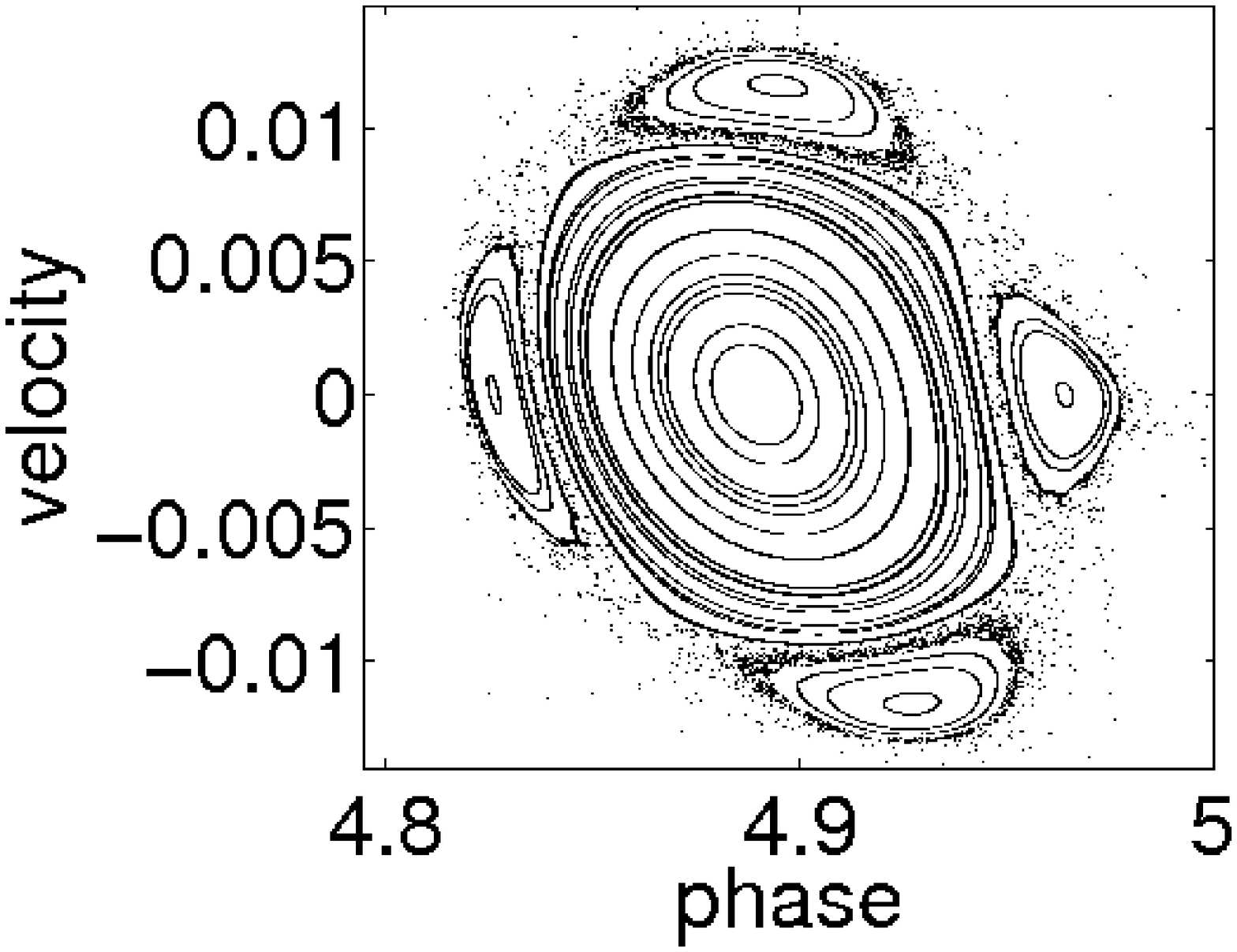}} \qquad
\subfigure[]{\includegraphics[width=0.22\textwidth]{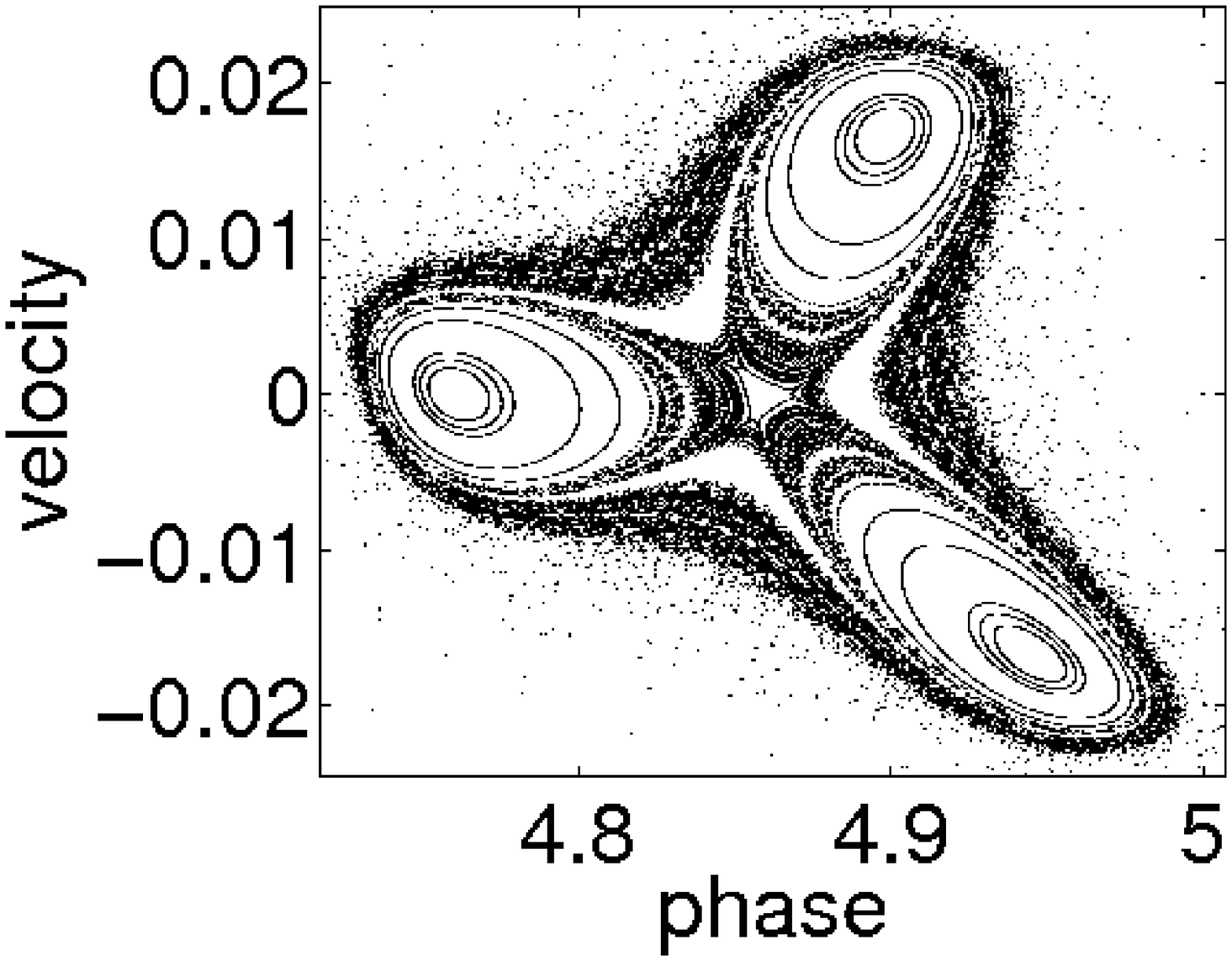}} \qquad
\subfigure[]{\includegraphics[width=0.22\textwidth]{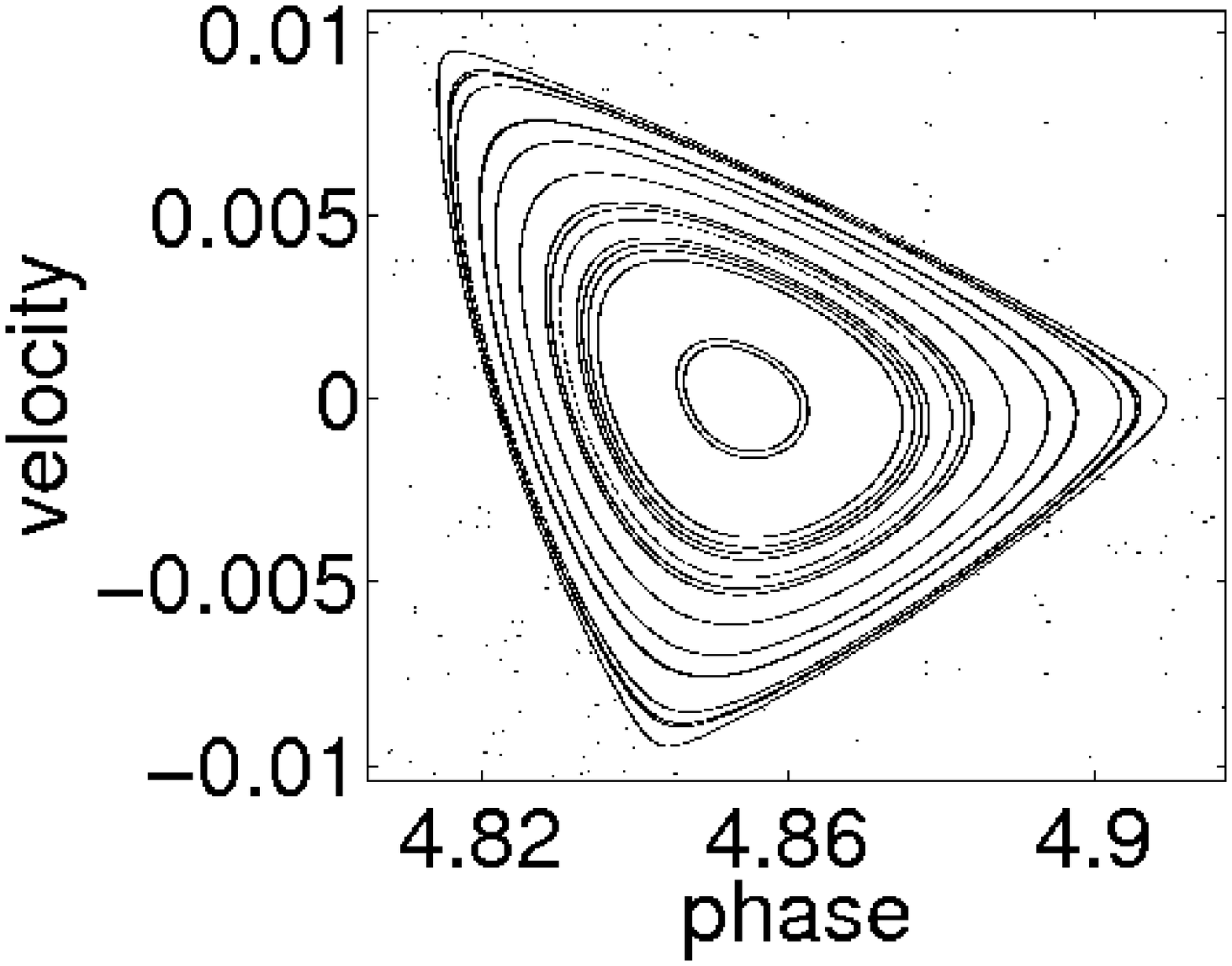}} \qquad
\subfigure[]{\includegraphics[width=0.22\textwidth]{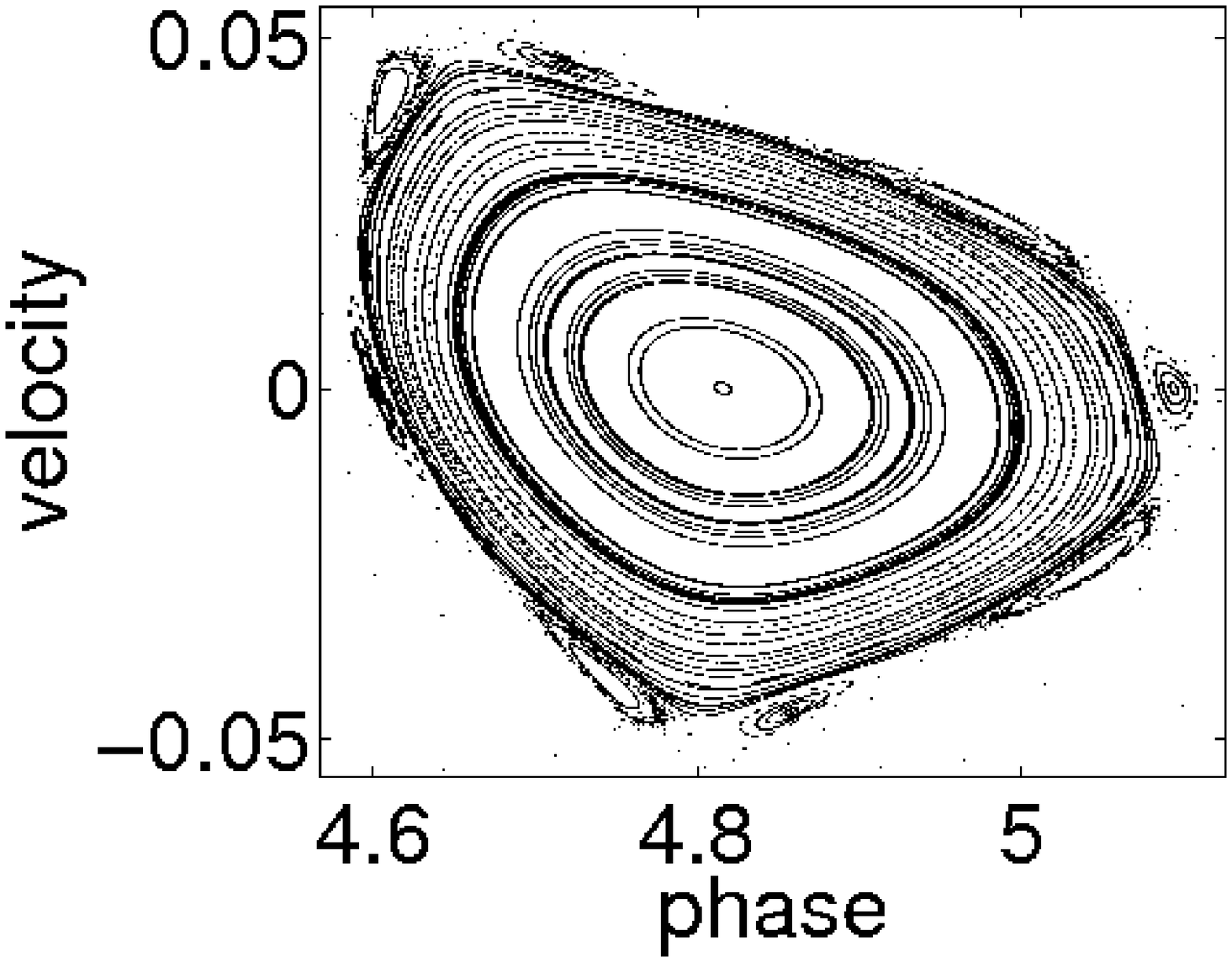}} \qquad
\subfigure[]{\includegraphics[width=0.22\textwidth]{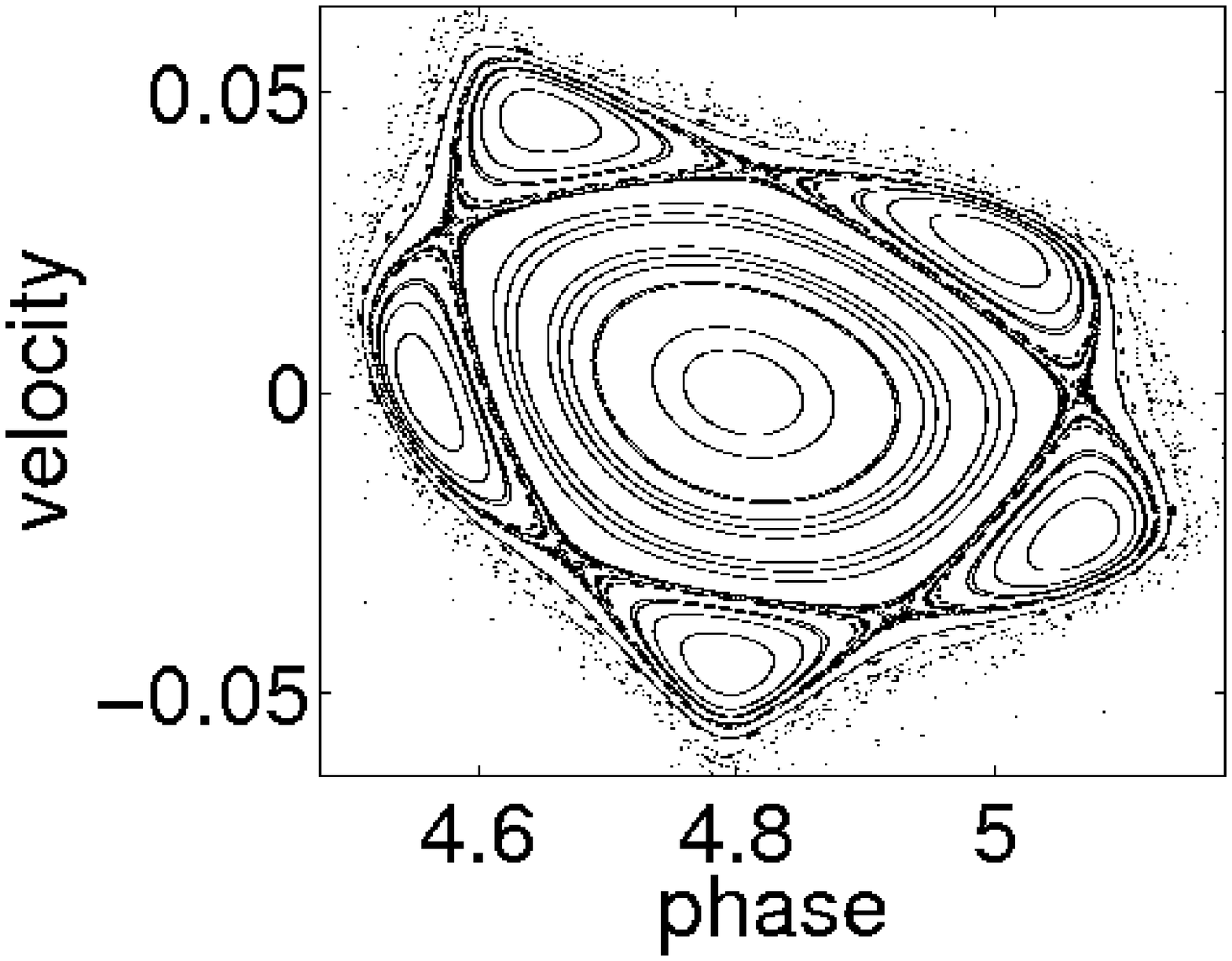}} \qquad
\subfigure[]{\includegraphics[width=0.22\textwidth]{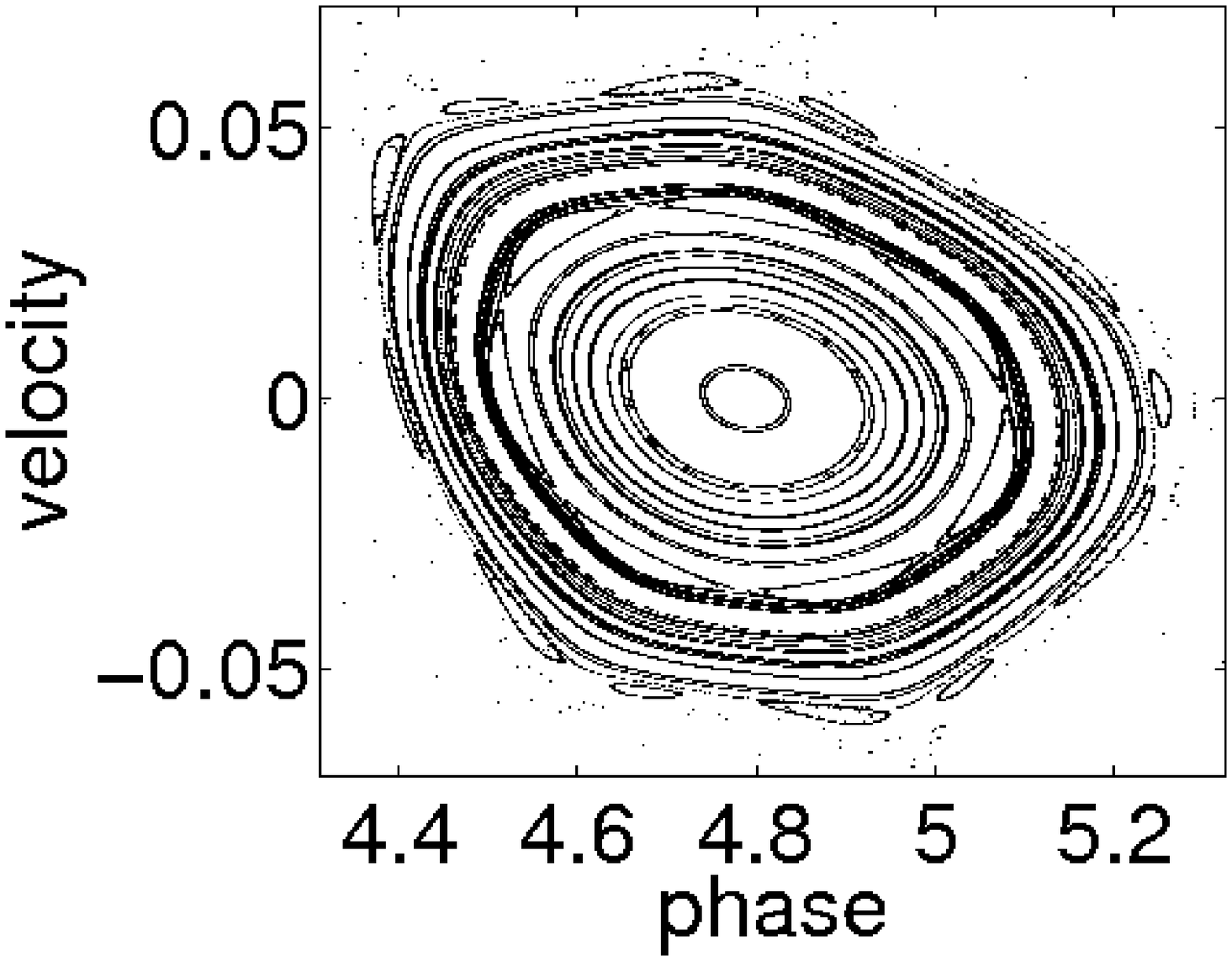}} \qquad
\subfigure[]{\includegraphics[width=0.22\textwidth]{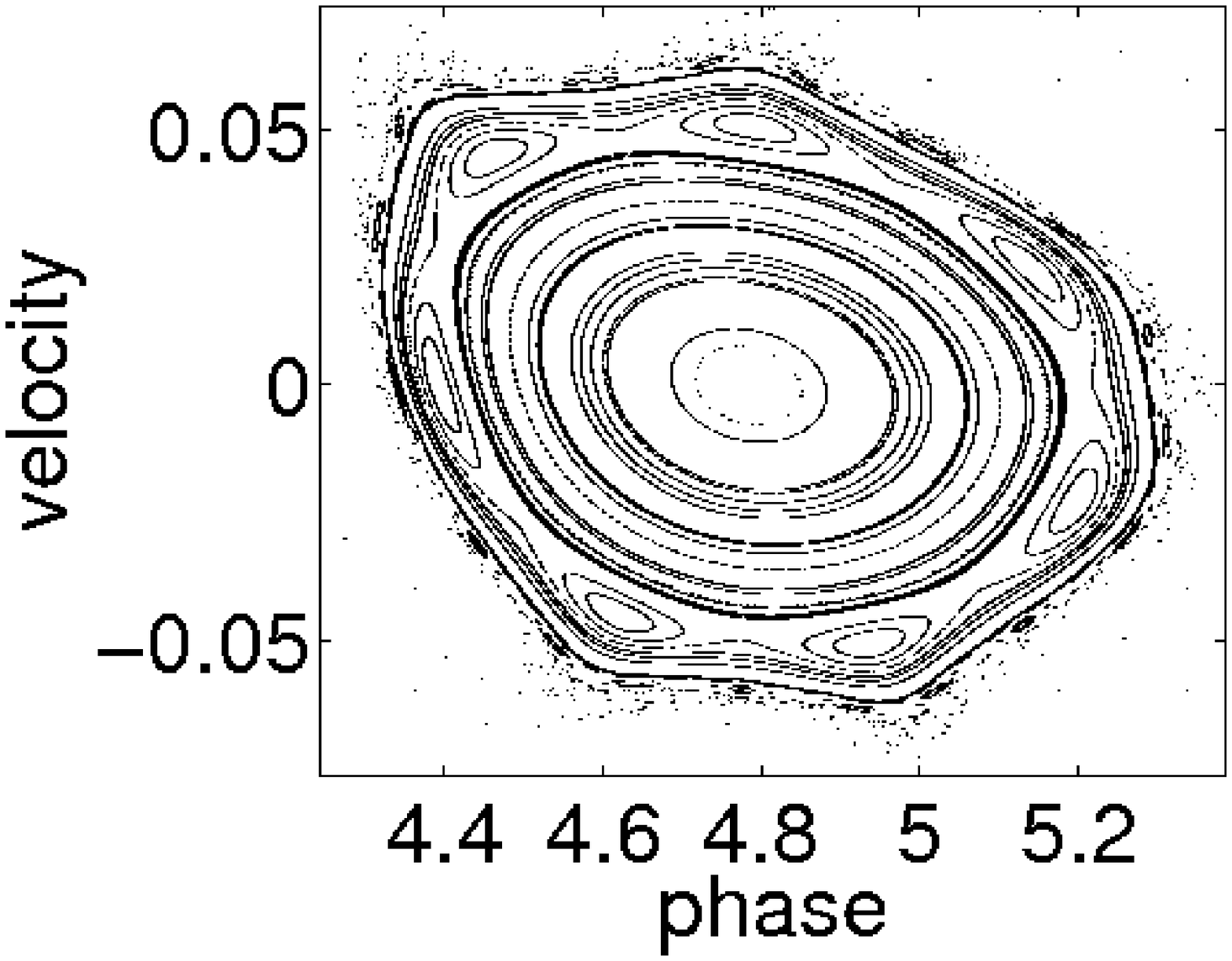}} \qquad
\subfigure[]{\includegraphics[width=0.22\textwidth]{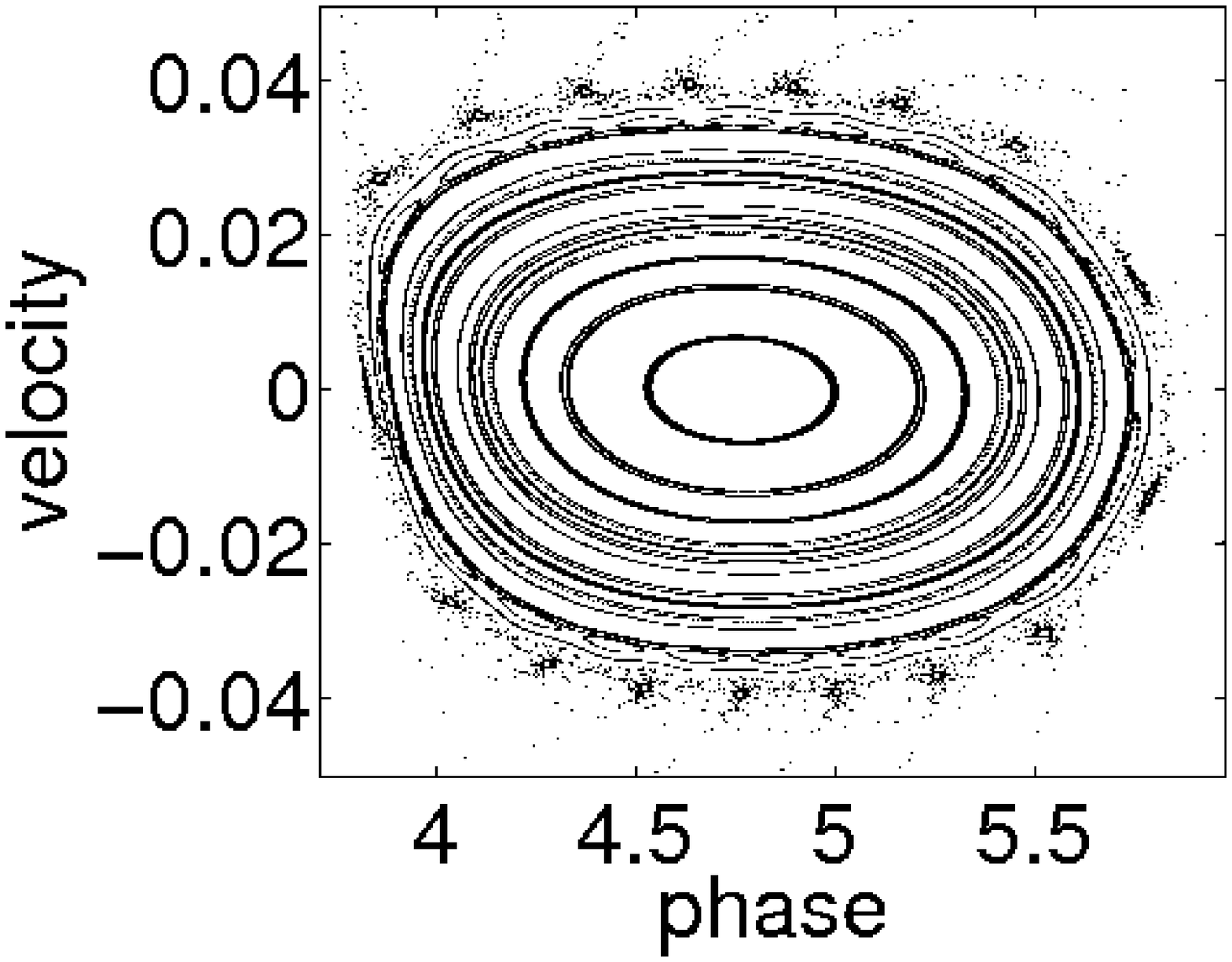}}
\caption{Overview of the parameter dependence of the KAM-island. The parameters are \(l=0.1\) and (a): \(V_0=0.895\) (b): \(V_0=0.867\) (c): \(V_0=0.86\) (d): \(V_0=0.76\) (e): \(V_0=0.7\) (f): \(V_0=0.553\) (g): \(V_0=0.5\) (h): \(V_0=0.05\)}
\label{fig_poincare_multi}
\end{figure}

The shape and size of the stable islands in the Poincar\'e sections change with varying parameters. Fig.~\ref{fig_poincare_zoom} is typical for the phase space of this system: The central fixed point and the island of quasi-periodic orbits are surrounded by a chaotic layer with a fractal structure of stable and unstable periodic orbits. Additionally, there exist one or more sets of large and distinguished sub-islands. These can be inside the main island, as the five sub-islands in Fig.~\ref{fig_poincare_zoom}, or outside of it. These sub-islands discern themselves from other KAM substructures in several ways. They are not only much larger than other sub-structures, but their creation and destruction follows a simple pattern as the parameters are changed. Keeping \(l\) constant while decreasing \(V_0\), the sub-islands of period \(n\) move towards the outer edge of the stable island. As \(V_0\) is decreased further, the sub-islands cross the outermost quasi-periodic orbit of the main island, forming separate sub-islands, and ultimately dissolve. Simultaneously, a new set of sub-islands of period \(n+1\) forms at the center of the stable region and begins to move to the outside. Islands with an even period appear as pairs, therefore the sequence of the number of large sub-islands is \(4, 3, 8, 5, 12, 7,...\), see Fig.~\ref{fig_poincare_multi}. The sequence starts at the maximal value of \(V_0\) that allows for bound orbits as plotted in Fig. \ref{fig_fix2}. In the limit of very small \(V_0\) the period \(n\) diverges and the sub-islands form almost a continuum that can not be resolved numerically. Between period \(n=4\) and \(n=3\) the KAM island takes a triangular form and its size goes to zero.
The creation and destruction of the sub-islands follows the same sequence for all \(l\) as \(V_0\) is varied, albeit on a different scale. A variation of the barrier width \(l\) at a constant \(V_0\) leads to the same sequence as well. This suggests that the qualitative behavior of the phase space structure can be described by just one effective nonlinearity parameter, consisting of a combination of \(l\) and \(V_0\).

This sequence of sub-islands is typical for nonlinear systems and has been studied in detail in the standard map, see e.g. \cite{Efthymiopoulos1997,Contopoulos1999}. The creation and destruction of stable periodic orbits, or rather quasi-periodic orbits in general, is tied closely to number theory. The KAM tori can be characterized by their winding number. It was conjectured \cite{Greene1979} that the last KAM tori to be destroyed when the nonlinearity is increased are those with rotation numbers equal to noble numbers, which can be written as continued fractions \footnote{A noble number is a irrational number the continued fraction of which has all elements equal to one, except for a finite number of elements. The most prominent noble number is the famous golden ration, which is also the most irrational number.}. Before such a torus with rotation number \(R\) is destroyed, all the periodic orbits with rotation number equal to a truncation of the noble number \(R\) become unstable. Thus, the winding numbers of the large sub-islands can be interpreted as the truncations of the simplest first-order noble numbers.

These sub-islands also play a significant role in the scattering process. As we will demonstrate, the flow into and out of the border zone of the stable island is dominated by the stable and unstable manifolds of the outermost large sub-islands, even for parameters for which the island has entirely dissolved.

\begin{figure}[htbp]
\includegraphics[width=0.48 \textwidth]{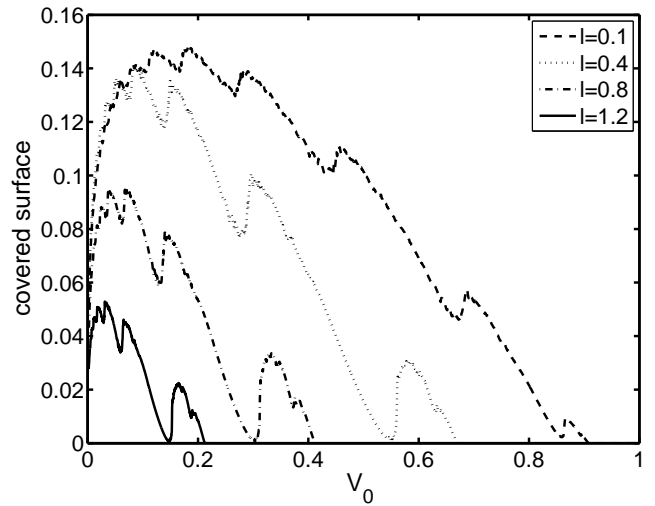}
\caption{Phase space volume of the stable island as a function of \(V_0\) for different \(l\).}
\label{fig_volume}
\end{figure}

The area covered by the elliptic island changes with the parameters as well. We analyzed numerically the surface covered by the stable island as a function of the parameters by dividing the phase space of the Poincar\'e sections into a fine grid of \(10^6\) small squares. All squares containing data points of the stable islands count as part of the surface. We checked this method for its stability by doubling the number of data points and comparing the resulting surfaces. The corresponding area is shown in Fig.~\ref{fig_volume} as a function of the barrier's height \(V_0\) for different values of the barrier's width \(l\). The peaks of the area in Fig.~\ref{fig_volume} are well understood: The size of the stable island is defined by two competing effects. As the potential hight \(V_0\) is decreased, the quasi-periodic orbits move outwards, which increases the island's size. At the same time, outer quasi-periodic orbits are destroyed. Since tori with noble rotation number are destroyed very late, the covered surface reaches a maximum whenever the outermost torus has a noble rotation number. At this point, other tori inside this noble torus have already been destroyed and the outermost curve separates a chaotic sea inside of the island from the outer chaotic area. When this outermost noble torus is destroyed, the two chaotic parts become connected and the size of the island reaches a sharp minimum. This kind of behavior is typical for nonlinear systems, see \cite{Contopoulos1999,Contopoulos2005}.

To gain further insight into the properties of the transition zone around the stable island, we use a method developed in \cite{Schmelcher1997} and \cite{Schmelcher1998} to locate the unstable periodic orbits (UPO) of this system. Given a N-dimensional discrete chaotic dynamical system \(U\) defined by
\begin{equation}
U: \vec{r}_{i+1}=\vec{f}(\vec{r}_i)
\end{equation}
a linear transformation is used to construct a new system \(S_k\) defined as
\begin{equation}
S_k: \vec{r}_{i+1}=\vec{r}_i+\lambda \mathbf{C}_k (\vec{f}^p(\vec{r}_i)-\vec{r}_i)
\end{equation}
where the matrix \(\mathbf{C}_k\) is orthogonal, \(\lambda\) is a small factor and \(p\) is the period of the fixed point we want to stabilize. Evidently, \(S_k\) and \(\vec{f}^p\) have the same fixed points. It can be shown that for every unstable fixed point there exists a suitable orthogonal transformation matrix \(\mathbf{C}_k\) and a small factor \(\lambda\) such that this unstable fixed point is stable under \(S_k\). The factor \(\lambda\) has to be small enough so that the eigenvalues of the matrix \(\mathbf{1}+\lambda \mathbf{C}_k (\mathbf{T}_U-\mathbf{1})\) have absolute values smaller than 1, where \(\mathbf{T}_U\) is the stability matrix of the system \(U\). The matrices \(\mathbf{C}_k\) correspond to reflections and rotations along the coordinate axes, thus all entries of \(\mathbf{C}_k\) are \(C_{ij}\in \{0,\pm 1\}\) and each row and column contains only one element which is different form zero. There exists a total number of \(N! 2^N\) of such matrices that will, in general, stabilize different types of unstable periodic orbits. However, it can be shown that a much smaller number of \(\mathbf{C}_k\) is sufficient to find all periodic orbits because some of the linear transformations are redundant. Only three matrices are needed in a two-dimensional system, see \cite{Pingel2004}. The advantage of this method over more conventional approaches, such as Newton-Raphson, is the global convergence. Even initial conditions far away from a fixed point eventually converge if the matrix \(\mathbf{C}_k\) and the factor \(\lambda\) are chosen appropriately.

\begin{figure}[htpb]
\includegraphics[width=0.48\textwidth]{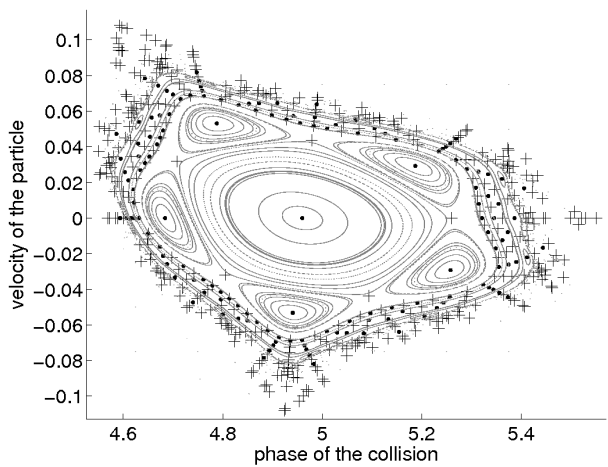}
\caption{Plot of the Poincar\'e section and the periodic orbits of period one to 21. Stable orbits are shown as dots, unstable orbits as crosses.}
\label{fig_UPO}
\end{figure}

In the system of the driven barrier we search for periodic orbits in the unique Poincar\'e sections such as Fig.~\ref{fig_poincare_zoom}. This way each point in the \((v,\varphi)\)-plane is uniquely connected to a point in the \((x,v,t)\) phase space. To find the periodic orbits we cover the \((v,\varphi)\)-plane with a grid of \(10^4\) initial conditions and iterate the transformed mapping with all three matrices \(\mathbf{C}_k\) and for periods of one to 21 with respect to the Poincar\'e section. The parameter \(\lambda\) is chosen between \(5 \cdot 10^{-3}\) and \(10^{-5}\) for higher periods. A typical result is plotted in Fig.~\ref{fig_UPO} which shows the unstable periodic points as black crosses, the stable periodic orbits as black dots and the corresponding Poincar\'e section in gray. The stability was derived according to the eigenvalues of the monodromy matrix. The chaotic layer around the elliptic island contains many families of periodic orbits which form the skeleton of the fractal structure. The number of unstable fixed points rises exponentially with their period.

\begin{figure}[htbp]
\subfigure[]{
\includegraphics[width=0.48\textwidth]{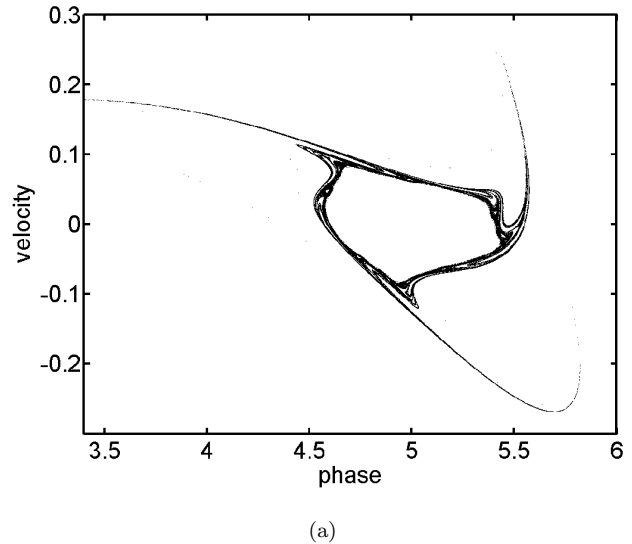}
%\caption{}
\label{fig_flowstable}}
\subfigure[]{
\includegraphics[width=0.48\textwidth]{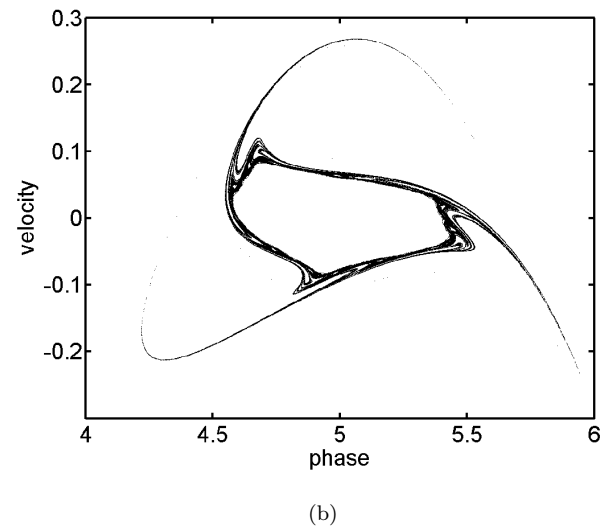}
%\caption{}
\label{fig_flowunstable}}
\caption{Stable (a) and unstable (b) manifolds of the UPOs of the system.}
\end{figure}

The unstable periodic orbits, or rather their stable manifolds, play an important role for the scattering process. This is because, although the stable island is itself not directly accessible by initial conditions starting outside of the interaction area, the stable manifolds, also called stable asymptotic curves, reach far into the part of phase space which is accessible from the outside. In order to calculate the flow of an unstable periodic orbit we take a small initial segment of length \(10^{-8}\) along the eigenvectors of the monodromy matrix at the UPO and iterate an ensemble of \(10^6\) initial conditions in this segment forward in time along the unstable asymptotic curves or backward in time along the stable asymptotic curves. It is in this system not possible to follow the manifolds of a specific UPO, because the manifolds of different UPOs cross each other at heteroclinic intersections. Thus the flow becomes chaotic and our simulations produce a global picture of the flow of the system of UPOs. The stable and unstable manifolds, plotted in Fig.~ \ref{fig_flowstable} and Fig.~\ref{fig_flowunstable}, reach far out of the border region into the chaotic sea in what looks in these plots as spiral arms, which consist of an infinite number of stable or unstable curves. These outer manifolds belong to periodic orbits with a period of 3 or multiples of 3.

We calculated the asymptotic curves of the system for the whole range of parameters and found that they are closely tied to the large primary sub-islands around the island. The flow into and out of the stable island is always dominated by the outermost family of sub-islands, even if this family is completely unstable. Therefore the number of spiral arms which are formed by the manifolds follows the same sequence as the sub-islands when the parameters are changed.
The unstable periodic orbits and with them their asymptotic curves are present only for parameters which allow for the existence of the stable island in phase space, see Fig. \ref{fig_fix2}.

\section{Scattering dynamics} \label{sec_scattering}

The phase space structure presented in section \ref{sec_phasespace} has profound effects on scattering processes. Due to the existence of a KAM-island in phase space, the ac-driven barrier is a chaotic scatterer. (See \cite{Ott1992} for a definition.) To simulate the scattering process we place an ensemble of particles with an uniformly distributed velocity \(v_{in}\) far outside of the scattering region. The initial phase \(\varphi_0\) (or time) is distributed in such a way that the phase of the first collision \(\varphi_1\) with the barrier is uniformly distributed in \([0, 2\pi]\). It makes sense to use the initial velocity \(v_{in}\) and the phase of the first collision \(\varphi_1\) as parameters of the scattering because this allows an easy comparison to the Poincar\'e sections of section \ref{sec_phasespace}, where we use similar coordinates. The disadvantage of these coordinates is that not all combinations of \(v_{in}\) and \(\varphi_1\) are accessible from the outside of the scattering region. Thus, the following plots have an area marked as inaccessible.

\begin{figure}[htbp]
\includegraphics[width=0.48\textwidth]{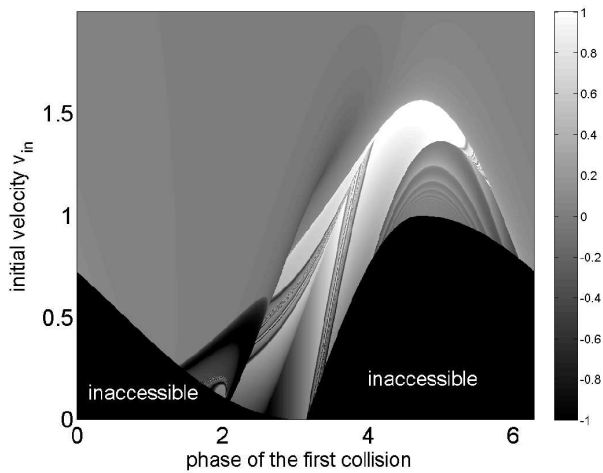}
\caption{Velocity change \(|v_{in}|-|v_{out}|\) of the particle as a function of velocity \(v_{in}\) and phase of the first collision \(\varphi_1\) in gray-scale.}
\label{fig_dv}
\end{figure}

As scattering functions we examine in detail the dwelltime, defined as the time the particles spend in the scattering region, and the change of the velocity of the particles, \(|v_{in}|-|v_{out}|\). We also determine the particles' number of collisions and whether the particles are transmitted or reflected. The velocity change is plotted in Fig.~\ref{fig_dv} for \(V_0=0.32\) and \(l=0.4\) in a gray-scale plot. Dark surface colors represent acceleration, bright areas stand for deceleration.  In the following, we analyze the scattering process for this representative set of parameters. It is possible to discern different well separated regions in Fig.~\ref{fig_dv} in which the scattering function is smooth. In other regions, most prominently in a wedge-shaped part in the center of Fig.~\ref{fig_dv}, the scattering function has unresolved parts. The scattering process is chaotic in these regions.

\begin{figure}[htbp]
\subfigure[]{\includegraphics[width=0.22\textwidth]{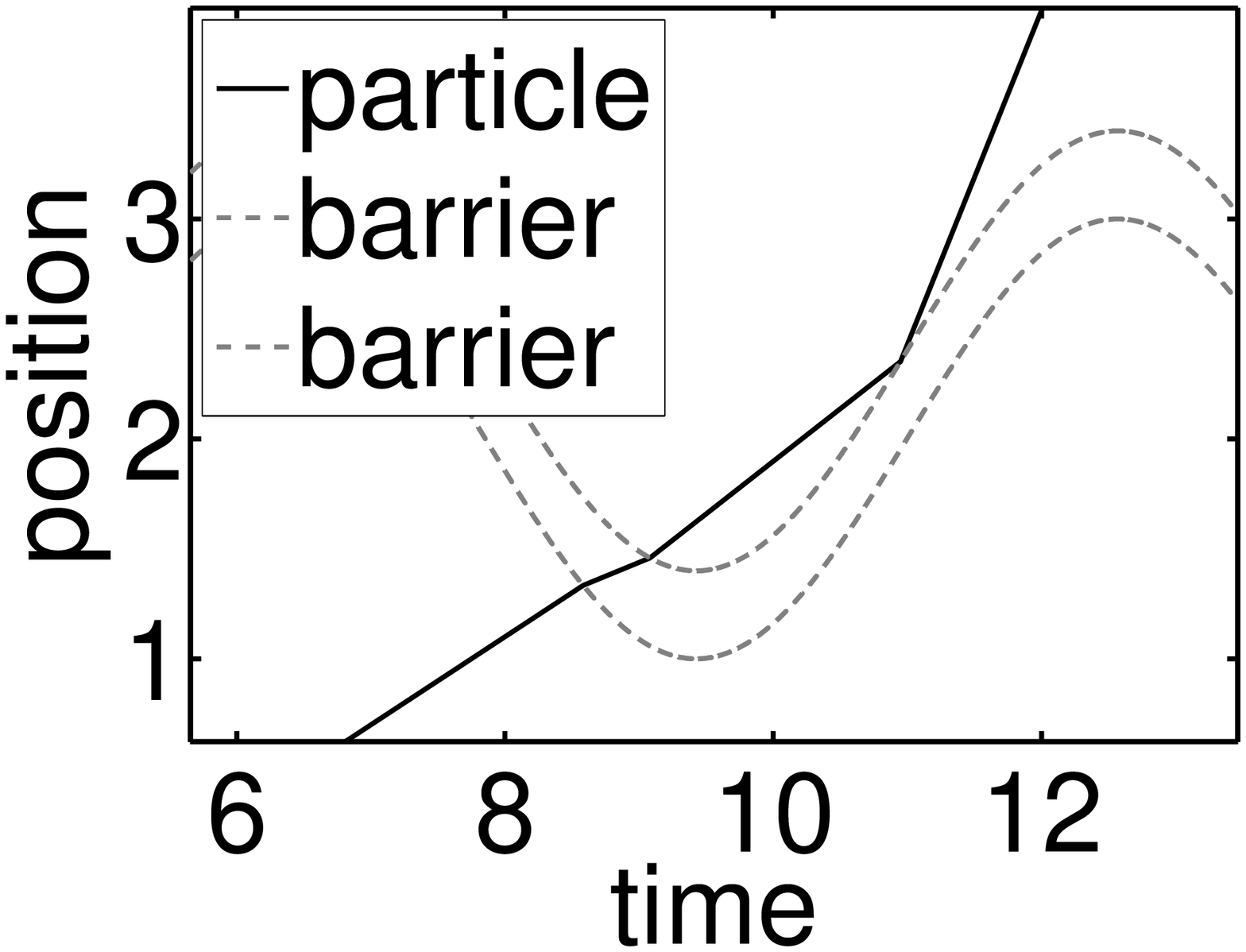} \label{fig_acceleration}}
\subfigure[]{\includegraphics[width=0.22\textwidth]{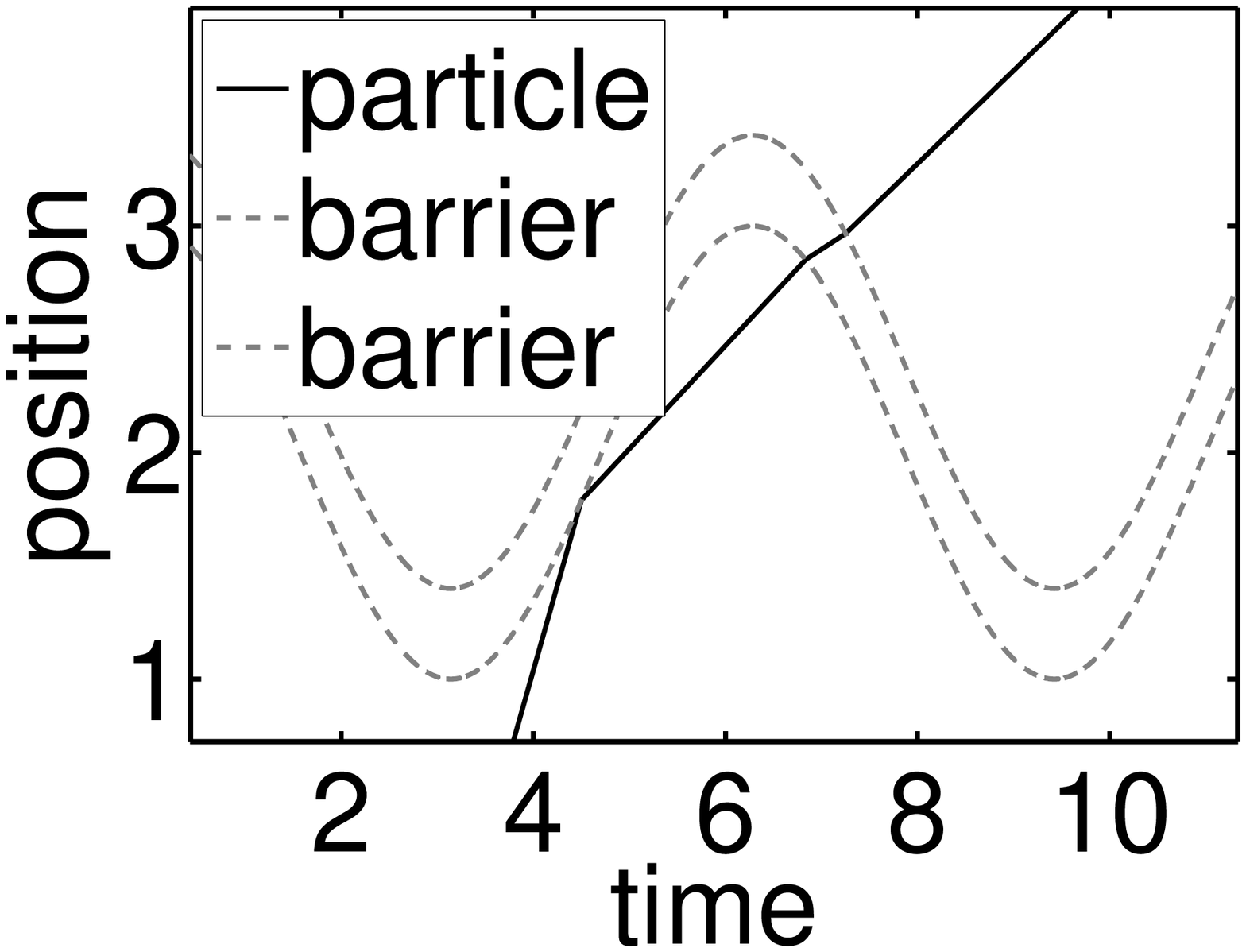} \label{fig_deceleration}}
\caption{Typical trajectories for acceleration (a) and deceleration (b) at (\(v_{in}=0.4\),\(\varphi_1=2.3\)) and (\(v_{in}=1.5\),\(\varphi_1=4.5\))}
\end{figure}

We discuss the case of regular scattering first: The velocity of fast particles is hardly changed at all, because fast particles simply traverse the barrier in a very short time. According to eq.~(\ref{equ_vn+1_a}), the transmission through the barrier is elastic if the velocity of the barrier at the collision with the left edge of the barrier is equal to the barrier velocity at the collision with the right edge, \(v_b(\varphi_1)=v_b(\varphi_2)\), which is approximately the case for fast particles. The scattering of slow particles is, in general, inelastic, particularly if a particle is reflected by the barrier, see eq.~(\ref{equ_vn+1_b}). Figs.~\ref{fig_acceleration} and \ref{fig_deceleration} show trajectories typical for acceleration and deceleration, respectively.

\begin{figure}[htbp]
\includegraphics[width=0.48\textwidth]{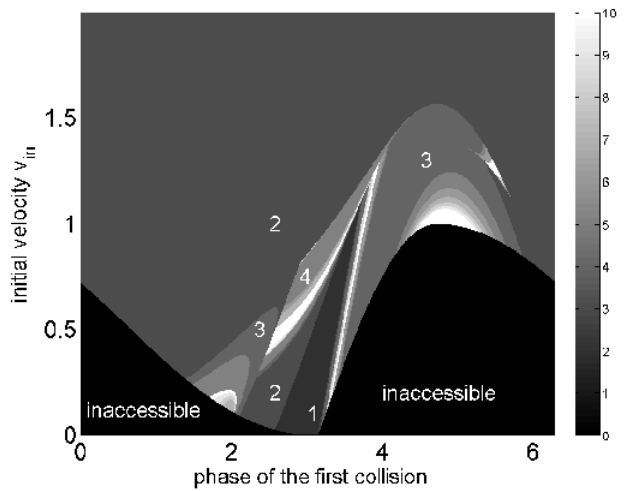}
\caption{Number of collisions of the particle as a function of velocity \(v_{in}\) and phase \(\varphi_1\) represented as gray-scale. The collision number is also printed in the plot.}
\label{fig_ncoll}
\end{figure}

The regular regions visible in Fig.~\ref{fig_dv} correspond to a constant number of collisions, as plotted in Fig.~\ref{fig_ncoll}. The edges of the regular regions correspond to a change in the number of collisions, not necessarily by one, which naturally leads to a sudden change in the other scattering functions. The number of collisions is small, between one to four,  for all regular scattering trajectories.

\begin{figure}[htbp]
\includegraphics[width=0.48\textwidth]{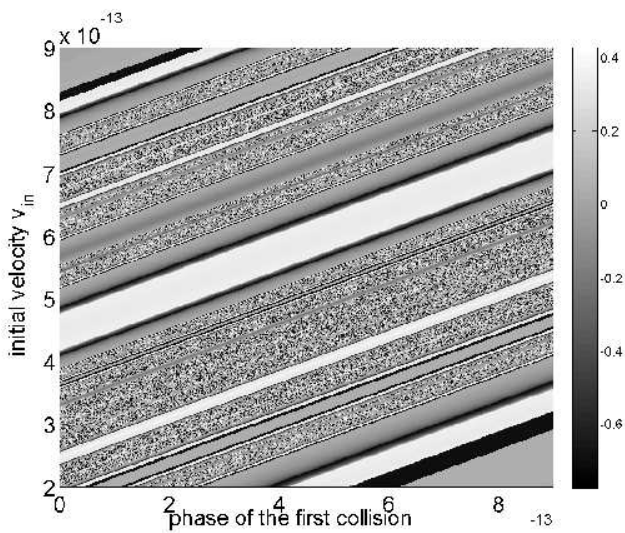}
\caption{Enlargement of part of Fig.~\ref{fig_dv} by a factor of \(10^{13}\).}
\label{fig_zoom}
\end{figure}

In the chaotic parts of the scattering function in Fig.~\ref{fig_dv}, the dynamics is infinitely sensitive on the initial conditions. Therefore it is not possible to resolve the scattering function in these areas completely. These unresolvable points are singularities. We have tested this by successively magnifying the irregular parts of Fig.~\ref{fig_dv} up to the numerical limits. Fig.~\ref{fig_zoom} shows such an enlargement by a factor of \(10^{13}\). The singularities of the scattering function, visible as unresolved parts in Fig.~\ref{fig_zoom} and \ref{fig_dv}, form a fractal set. Actually, the chaotic parts of the scattering function are predominant on smaller scales, i.e. the smooth parts, still visible in Fig.~ \ref{fig_zoom}, become rare. This is typical for non-hyperbolic chaotic scattering (\cite{Lau1991}).

\begin{figure}[htbp]
\includegraphics[width=0.48\textwidth]{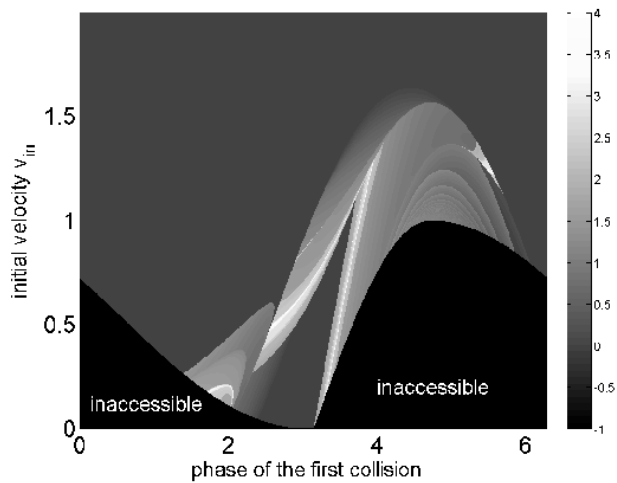}
\caption{Logarithm of the dwelltime of the particle as a function of velocity \(v_{in}\) and phase \(\varphi_1\) represented as gray-scale.}
\label{fig_dwelltime}
\end{figure}

The origin of the chaotic scattering is the stickiness of particles to the KAM-island in phase space. This can be seen from a logarithmic plot of the dwelltime in Fig.~\ref{fig_dwelltime}. The dwelltime diverges in exactly the regions in which the scattering function has singularities. The number of collisions diverges as well in these chaotic regions. (This can not really be seen in Fig.~\ref{fig_ncoll}, because the color map is capped at \(n=10\). But the collision number reaches  \(n=10^4\) and more in the white parts of the plot.) Trajectories starting on initial conditions in the chaotic parts of the scattering function become sticky and trace the outermost quasi-periodic orbits of the stable island for arbitrary long times. The number of collisions per time unit is therefore constant for all sticky trajectories, namely four collisions per period \(T\) of the driving function.

It is known that in most systems which exhibit chaotic scattering, singularities in the scattering function have a divergent dwelltime. (\cite{Ott1992}) This can be easily understood since the scattering function is infinitely sensitive to perturbations of the initial conditions leading to singularities which can only be the case if the interaction time in continuous systems or the number of interactions in discrete systems between target and particle diverges as well.

\begin{figure}[htbp]
\includegraphics[width=0.48\textwidth]{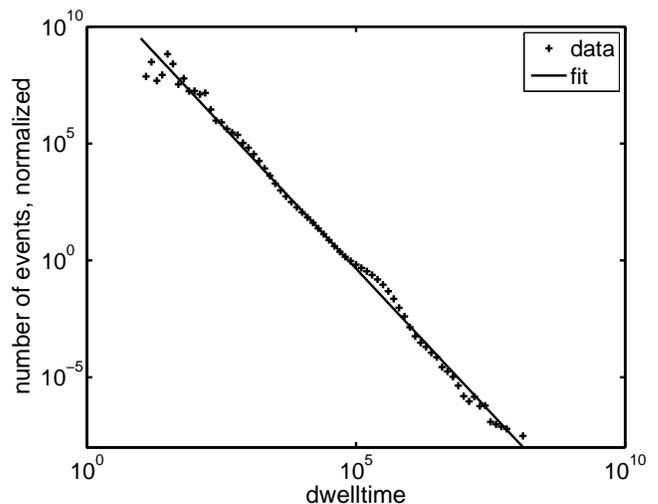}
\caption{The distribution of the sticking times, using a logarithmic adjusted bin size. By fitting a power law we found an exponent of \(\gamma=-2.5\).}
\label{fig_sticking_times}
\end{figure}

The dwelltimes of sticky particles have a typical probability distribution. Fig.~\ref{fig_sticking_times} shows the distribution of dwelltimes in the sticky regions of Fig.~\ref{fig_dwelltime} for a total of more than \(10^{10}\) random initial conditions. The dwelltime distribution can be approximated by a power law \(P(t_d)\sim t_d^{\gamma}\). We find an exponent of \(\gamma=-2.5\) from a fit to Fig.~\ref{fig_sticking_times} between \(t_d=10^1\) and \(t_d=10^8\).

The origin of the stickiness in this system are the stable manifolds of the unstable periodic orbits described in section \ref{sec_phasespace}. Although the KAM-island itself is inaccessible by scattering trajectories, the stable manifolds reach out of the inaccessible area and intersect the set of scattering initial conditions. These intersections are identical to the singularities of the scattering function and have an infinite time delay. The fractal structure of the scattering function in the chaotic regions is just the structure of the stable manifolds, which are formed by the unstable periodic orbits in the KAM-island.

Another way of characterizing a chaotic scattering process is to calculate the uncertainty dimension of the chaotic part of the scattering function, which is a proxy for the fractal dimension. The uncertainty dimension describes the scaling with the resolution of the proportion of singular points to regular points in a two-dimensional scattering function. As predicted in \cite{Lau1991} for nonhyperbolic chaotic scattering systems, such as the ac-driven barrier, the uncertainty dimension is exactly one for all parameters that allow for the existence of the stable KAM island.

\begin{figure}[htbp]
\subfigure[]{\includegraphics[width=0.22\textwidth]{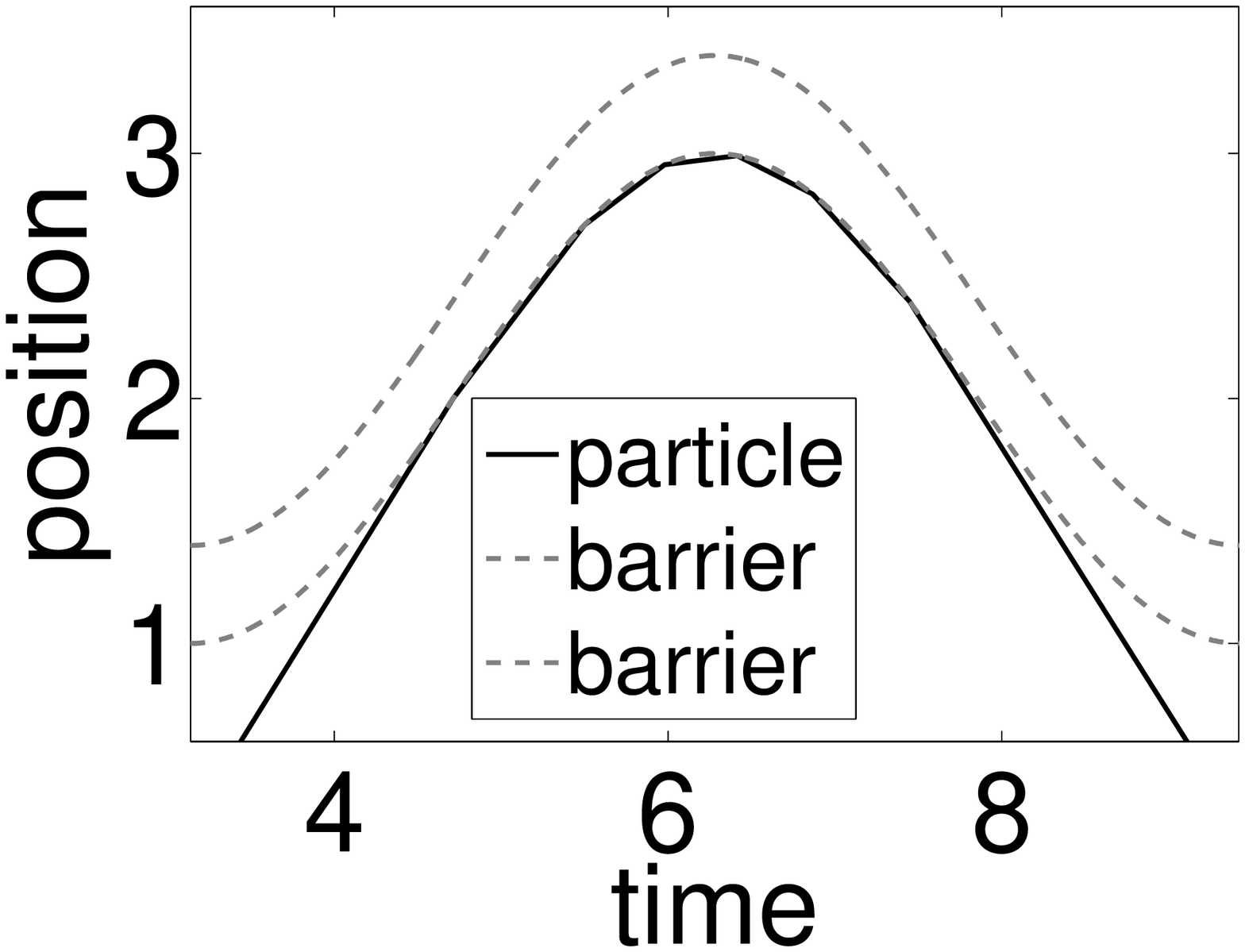} \label{fig_whisper}}
\subfigure[]{\includegraphics[width=0.22\textwidth]{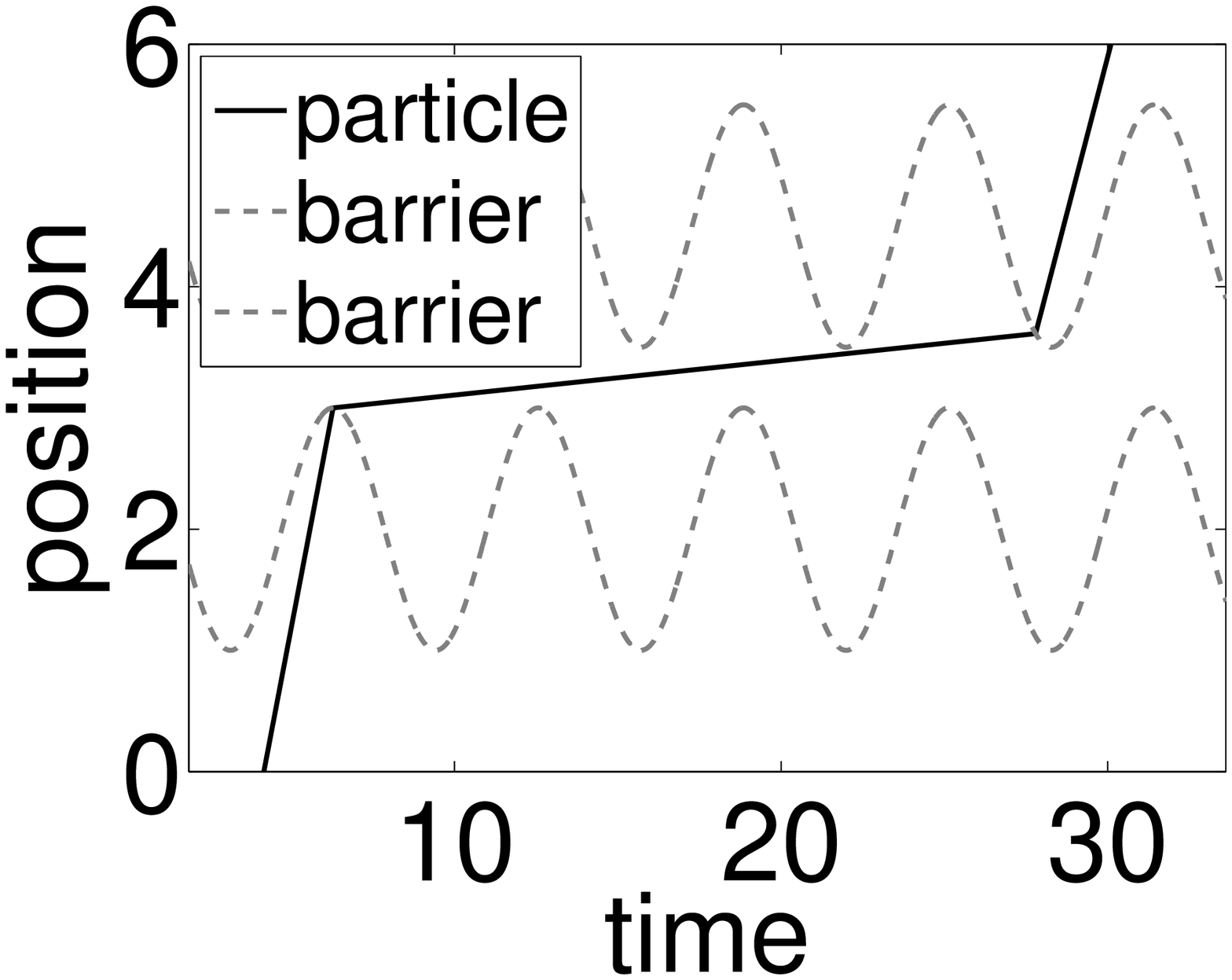} \label{fig_LVP}}
\caption{Singularities of the scattering function not associated with chaotic scattering. (a) is the trajectory of a whispering gallery orbit, (b) is a low velocity peak.}
\end{figure}

The scattering function has two other singularities, which are not caused by stickiness to regular structures. Such singularities are isolated and do not lead to chaotic scattering. Nevertheless, the scattering function is infinitely sensitive to changes in the initial conditions at these singularities. One of them is located at \(v_{in}=1\) and \(\varphi_1=1.5\pi\) in Fig.~\ref{fig_dv} and is called ``whispering gallery'' in static systems. The dwelltime, see Fig.~\ref{fig_dwelltime}, is quite small at this point, \(t_d=T/2\), whereas the number of collisions diverges, see Fig.~\ref{fig_ncoll}. A particle on this trajectory hits the barrier at \(\varphi_1=1.5\pi\) with a velocity slightly larger than 1, \(v_{in}=1+\epsilon\). The barrier has its maximum velocity of \(v_b=1\) at this phase. The particle is reflected and the new velocity becomes \(v_1=2 v_b-v_{in}=1-\epsilon\) according to eq.~(\ref{equ_vn+1_b}). Because the barrier decelerates after \(\varphi=1.5\pi\) the particle collides with the barrier again after a very short time at \(\varphi_2\approx1.5\pi+\sqrt{6\epsilon}\). The particle is decelerated by this collision further to \(v_2\approx2 v_b-v_1=2(1-3\epsilon)-(1-\epsilon)=1-5\epsilon\). This sequence of successive collisions continues until the barrier reaches the turning point at \(\varphi=2\pi\) from where, due to the symmetry of the driving function, the process is inverted and the particle is accelerated by successive collisions. Since the particle has the last collision at \(\varphi=0.5\pi\) the dwelltime is only \(t_d=T/2\). However, for sufficiently small \(\epsilon\) the particle can have an arbitrarily large number of collisions, because the time passed between the collision is proportional to \(\epsilon\). A typical trajectory is plotted in Fig.~\ref{fig_whisper}.

The other singularity is a so called low velocity peak, \cite{Papachristou2001}. It only appears if the width of the barrier is larger than twice the oscillation amplitude, \(l>2\), and if the potential height is greater than the effective potential, \(V_0>1\). For these parameters, a particle can hit the barrier at the extremal position in \(\varphi=0\) at a velocity of exactly \(v_{0}=\sqrt{V_0}+\epsilon\). Such a particle is transmitted into the barrier and is decelerated to the small velocity \(v_1=\sqrt{\epsilon}\). Due to the large width of the barrier, the dwelltime of the particle becomes approximately \(t_d\approx\frac{l-2}{\sqrt{\epsilon}}\) and thus diverges. A typical trajectory of a low velocity peak is plotted in Fig.~\ref{fig_LVP}.

As the parameters of the system are changed, the regular parts of the scattering function (Fig.~\ref{fig_dv}) are deformed and shifted to different initial velocities, but remain qualitatively similar. The shape and position of the chaotic parts stay qualitatively similar as well, whereas their exact structure depends critically on the parameters. This is because the chaotic regions are created by the stable manifolds of the UPOs. The shape of the stable manifolds is directly connected to the primary sub-islands and therefore follows the sequence of creation and destruction described in section~\ref{sec_phasespace} as the parameters are changed.
Since the stable island and the surrounding unstable periodic orbits are the cause of the chaotic scattering, the scattering on the ac-driven barrier becomes regular for parameters that don't allow for elliptic orbits in phase space (see Fig.~\ref{fig_fix2}).
The singularity corresponding to the whispering gallery is independent of the parameters of the system, it exists in all harmonically laterally driven systems and it is not a unique property of the driven barrier. The low velocity peaks require a large barrier height \(V_0>1\) and width \(l>2\). Consequently, there are no parameters for which the low velocity peaks and the chaotic scattering due to the KAM-island coexist. It was shown in Refs. \cite{Papachristou2002,Papachristou2004} that low velocity peaks can lead to a new form of scattering dynamics, called dilute chaos, which requires the existence of UPOs and therefore does not appear in this system.

\subsection{Application and comparison to the quantum behavior} \label{subsec_QM}

\begin{figure}[htbp]
\subfigure[]{\includegraphics[width=0.22\textwidth]{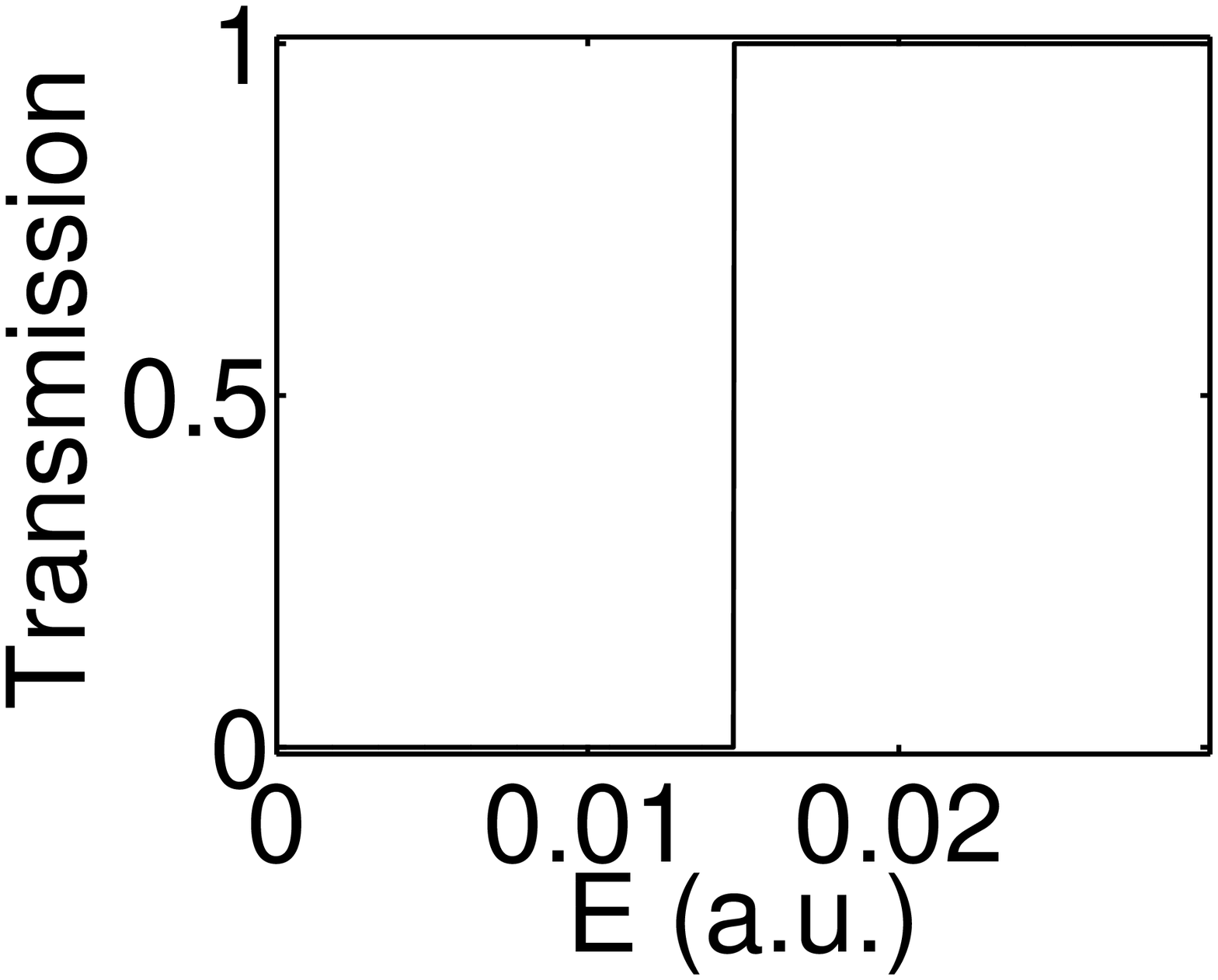}}
\subfigure[]{\includegraphics[width=0.22\textwidth]{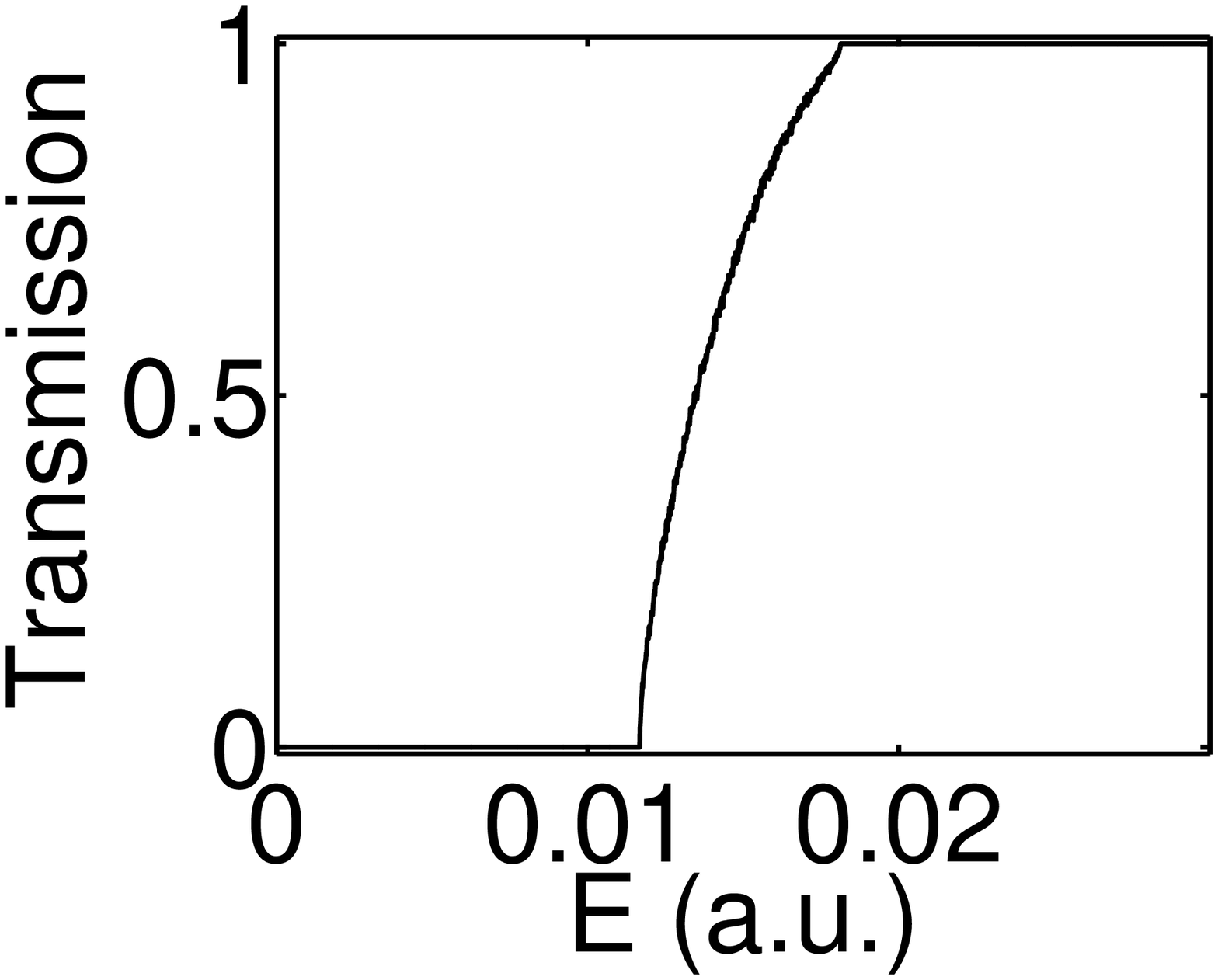}}
\subfigure[]{\includegraphics[width=0.44\textwidth]{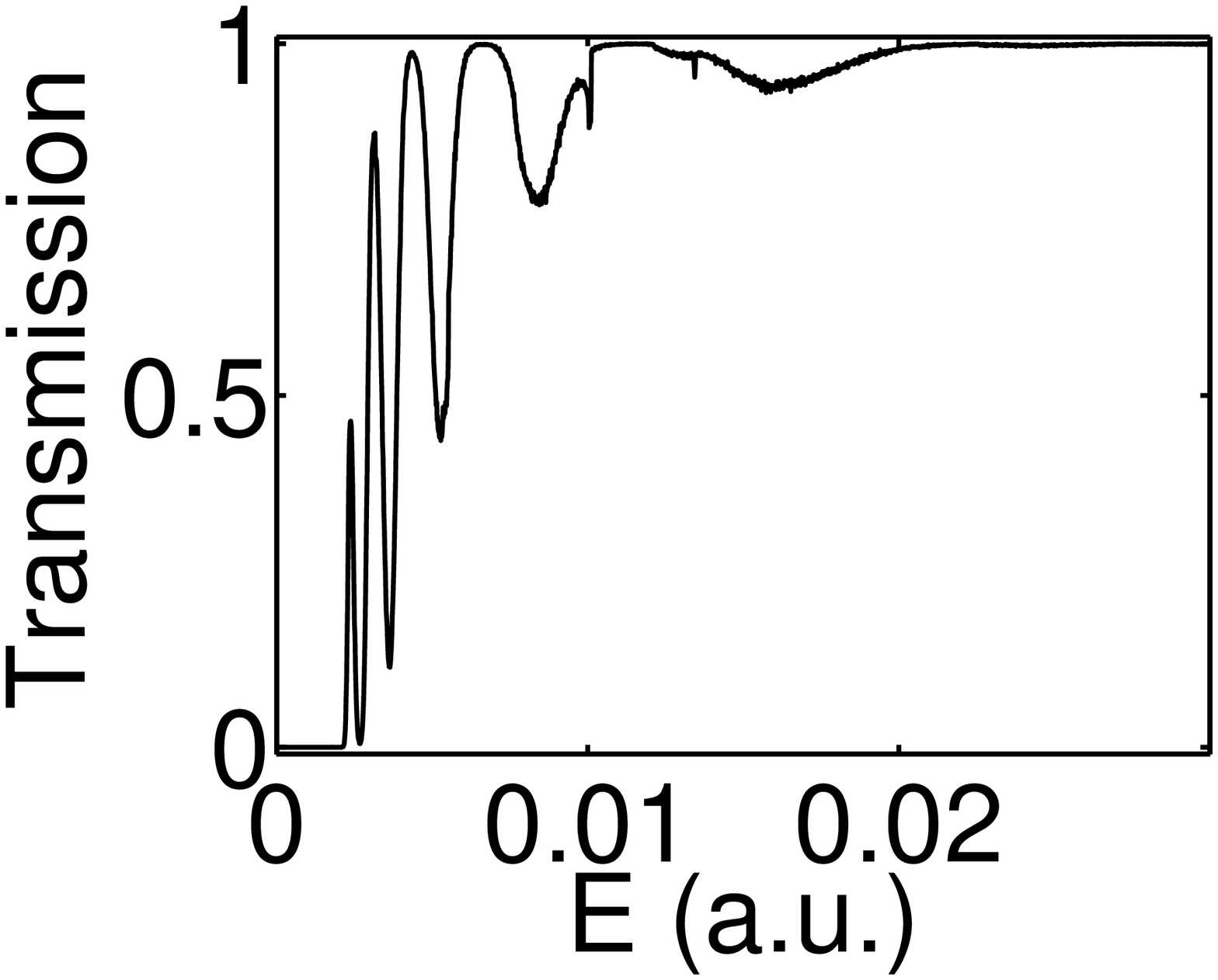} \label{fig_transmission_packet_2}}
\caption{Transmission function for different driving frequencies of the classical system. (a): \(\omega=0\), (b): \(\omega=0.0003\), (c): \(\omega=0.03\)}
\label{fig_transmission_packet}
\end{figure}

We can use our knowledge of the system to make comparisons with established results for the dynamics of the driven barrier in the quantum regime. The tunneling through a periodically driven square potential barrier has been analyzed in \cite{Vorobeichik1998}. (See also \cite{Chiofalo2003}.) It was found that the transmission coefficient as a function of the particle energy has resonances below the minimal tunneling energy of the static system for high driving frequencies. This can be explained by resonant tunneling into semi-stable bound states of an effective time-averaged potential, which has a double-barrier structure. We now want to compare the results in the quantum regime with our classical simulations. The parameters chosen in \cite{Vorobeichik1998} are \(a_0=200\), \(l=80\), \(m=0.1\) and \(V_0=0.0147\). The system is studied for three values of the driving frequency, \(\omega=0\), \(\omega=3\cdot10^{-4}\) and \(\omega=3\cdot10^{-2}\). The corresponding effective parameters are \(l=0.4\) and \(V_0\rightarrow\infty\), \(V_0=81.6\) and \(V_0=0.0082\). To imitate the transmission of a quantum wave packet with a classical simulation we use an ensemble of particles that is distributed as a minimal uncertainty Gaussian in position and momentum space, i.e. \(\sigma_x \sigma_p=\frac{1}{2}\). The results of our classical simulation are shown in Fig.~\ref{fig_transmission_packet}.

Surprisingly, the classical simulation coincides to an amazing degree with the results of the quantum mechanical analysis in the high frequency limit (Fig.~\ref{fig_transmission_packet_2}). The transmission in the classical system shows the same resonant behavior as in the quantum regime. (It should be mentioned that a fine tuning of the initial conditions is necessary to reproduce the resonances of \cite{Vorobeichik1998} exactly.)
Since the model of the effective time-averaged potential \(V_{eff}\) fails in the classical regime, this result is surprising. The scattering of classical particles off the static potential \(V_{eff}\) would simply reproduce a step function, since the classical mechanics just does not allow any tunneling into resonant states. It is important to note that the effective potential is generally ill suited to describe the dynamics of trapped particles in the classical regime. Although the effective potential could in principle be used to explain the existence of trapped particles for high frequencies, we also find trapped particles for low frequencies, i.e. when the driving frequency and the oscillation frequency of the trapped particles are of similar order of magnitude. The trapping is not caused by an effective potential for high driving frequencies but by a synchronization of the motion of the particle and the barrier.

\begin{figure}[htbp]
\subfigure[]{\includegraphics[width=0.44\textwidth]{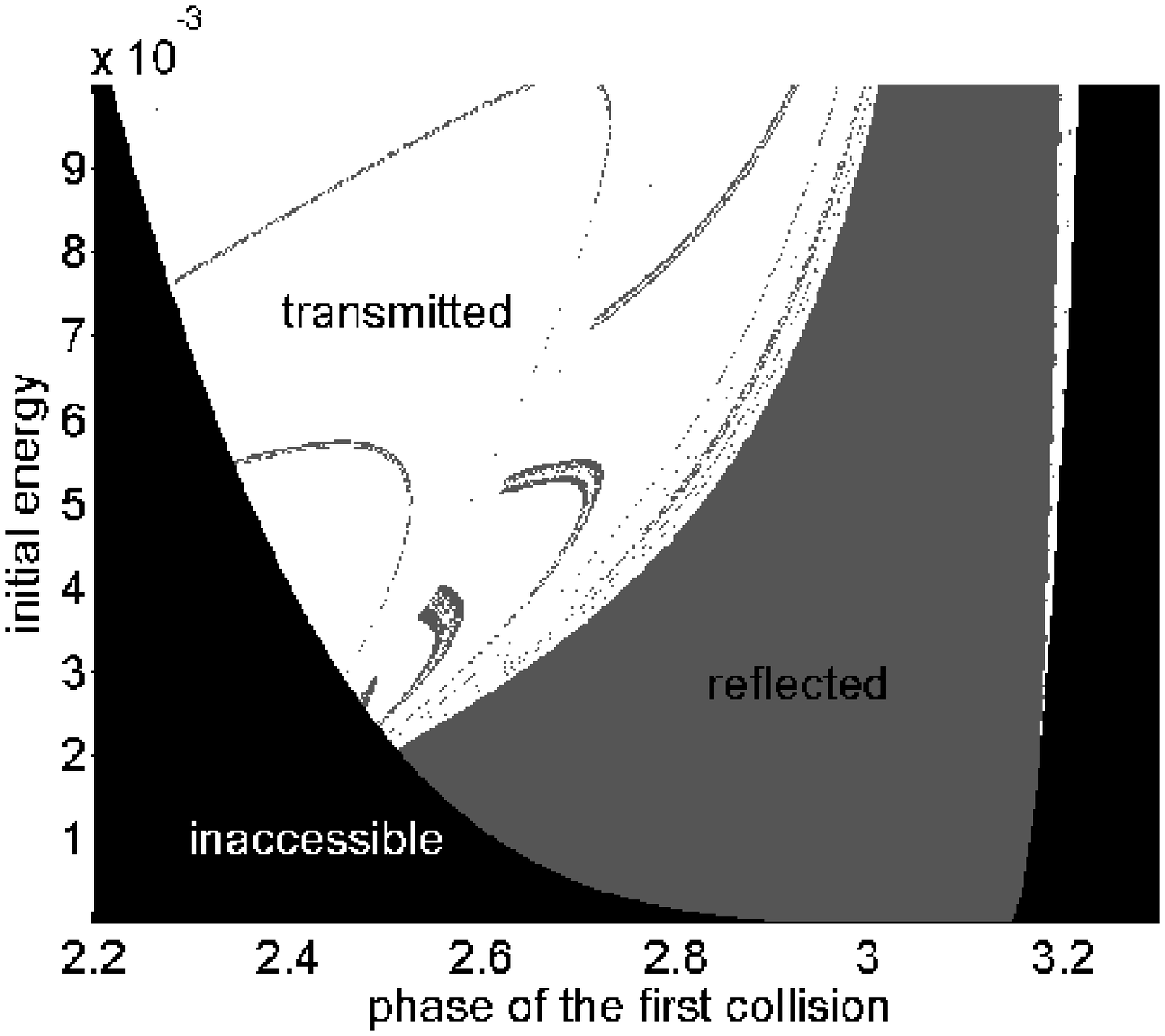} \label{fig_trans_fi1}}
\subfigure[]{\includegraphics[width=0.44\textwidth]{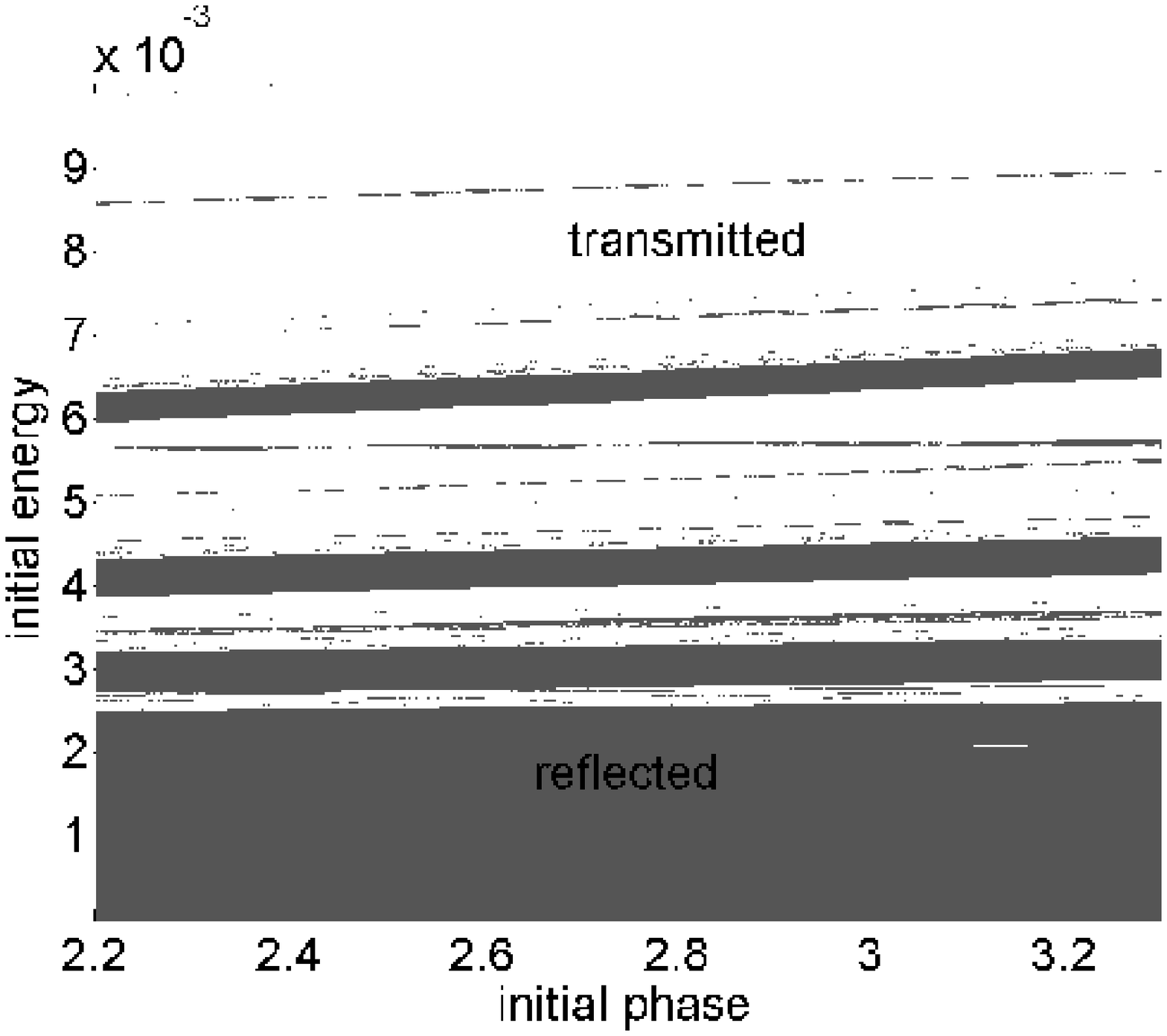} \label{fig_trans_fi0}}
\caption{Transmission function of a single particle as a function of the initial velocity and (a): the phase of the first collision \(\varphi_1\) (b): the initial phase \(\varphi_0\).}
\end{figure}

The origin of the resonances of the transmission function in the classical regime is very different from the quantum regime. The transmission function of a single particle in Fig.~\ref{fig_trans_fi1} does not show any resonances as a function of the initial energy, at least not of the form seen in Fig.~\ref{fig_transmission_packet_2}. (It makes sense to compare the transmission function of a single particle Fig.~\ref{fig_trans_fi1} with Fig.~\ref{fig_transmission_packet_2}, because using a narrowly distributed ensemble changes the qualitative behavior very little.) The reason is that the phase of the first collision \(\varphi_1\), used in Fig.~\ref{fig_trans_fi1} as coordinate, is not an appropriate coordinate to describe this scattering process. In the simulation leading to the transmission resonances in Fig.~\ref{fig_transmission_packet_2} the initial phase \(\varphi_0\) is kept constant whereas \(\varphi_1\) oscillates wildly as a function of the energy \(E\). When the initial velocity \(v_{in}\) of a particle starting at a distance of \(x_0\) from the scattering region is changed by a small amount \(\Delta v\), the time when the particle enters the scattering region is changed by \(\Delta t=\frac{x_0}{v_{in}^2} \Delta v\). In all these simulations, the initial velocity \(v_{in}\) is very slow, therefore the variation of the initial energy \(E\) leads to a fast oscillation of the collision phase \(\varphi_1\). When we plot the transmission as a function of the initial phase in Fig.~\ref{fig_trans_fi0}, the resonances of Fig.~\ref{fig_transmission_packet_2}  become visible.

Whether an incoming particle is transmitted or reflected depends, for small particle energies, mostly on whether the barrier is approaching the particle or receding from it at the time when the collision occurs. A variation of the energy for a constant collision phase \(\varphi_1\) has little impact on the transmission probability, see Fig.~\ref{fig_trans_fi1}. Therefore, these resonances are not really resonances of the \emph{energy} of the particle, as in the quantum regime. They are produced simply by the \emph{propagation} to the barrier, where the different collision phases lead to peaks in the transmission probability.

It would be very interesting to analyze the similarity between the classical and the quantum case in the framework of semiclassical physics. Doing so would require applying the Gutzwiller formula to the periodic orbits of the system to derive a semiclassical propagator for the system, see \cite{Brack1996}. However, such a study is beyond the scope of this work. It should be noted that in related systems such as an oscillating hard wall potential, semiclassical approximations did not work \cite{Cheng1993,Luz1994}.

\section{Summary and outlook} \label{sec_summary}

The aim of this work is to analyze and understand in detail the classical dynamics of the ac-driven barrier for the full range of parameters. Although the potential is repulsive, the system exhibits a dynamical trapping process which is associated with an island of stability. This trapping process can be understood as a synchronization process between particle and barrier, which depends on the curvature of the driving law. The stable KAM-island of quasi-periodic orbits in phase space leads to topological chaos. The central periodic orbit, and with it all stable and unstable periodic orbits, exist only for a limited range of parameters. We determined these parameter ranges, calculated the position of the period four orbit, the size of the elliptic island and its shape as a function of the parameters. The transition zone around the stable island contains an infinite set of unstable periodic orbits, the stable manifolds of which reach far away from the stable island. These stable manifolds make the system a chaotic scatterer. Initial conditions starting on the stable manifolds are singularities and have a divergent dwelltime and collision number. The singularities form a fractal set with an uncertainty dimension of one. The system possesses two additional types of singularities, the whispering gallery and a low velocity peak. These are isolated singularities and are not connected to the KAM-structure in phase space.
The transmission function of a suitably prepared ensemble yields results which are very similar to tunneling resonances in the quantum mechanical regime. However, the origin of these resonances is very different in the classical regime and this sheds a new light on the high frequency behavior of the driven barrier.

The results of this work all depend on the existence of a dynamical trapping process. The stable orbits which we discovered rely on the curvature of the harmonic driving law. When we use a sawtooth shaped driving law, we find no such stable orbits. It can be assumed that other suitably curved driving functions lead to bounded motion as well. Likewise, our results do not depend on the exact shape of the barrier itself.

\section{Acknowledgment}

Financial support by the FRONTIER program of the University of Heidelberg in the framework of the excellence initiative of the federal and state governments is gratefully acknowledged.

\end{document}